\documentclass[twoside]{article}
\usepackage[accepted]{aistats2020}

\usepackage[utf8]{inputenc}
\usepackage[T1]{fontenc}
\usepackage{lmodern}
\usepackage{hyperref}       
\usepackage{booktabs}
\usepackage{amsfonts}
\usepackage{amsmath}
\usepackage{nicefrac}
\usepackage{microtype}
\usepackage{color, soul}
\usepackage{tikz,pgf}
\usetikzlibrary{positioning}
\usepackage{float}
\usepackage{subfig}
\usepackage{xr}
\usepackage{multirow}
\usepackage{array}
\usepackage{standalone}
\usepackage{mathrsfs}
\usepackage{enumitem}
\usepackage{algorithm}
\usepackage[noend]{algpseudocode}

\usepackage{lineno}

\usepackage[round]{natbib}

\def\ci{\perp\!\!\!\perp}

\newenvironment{thma}[1]{\par\noindent{\bf Theorem #1\ }\em}{\em}
\newenvironment{lema}[1]{\par\noindent{\bf Lemma #1\ }\em}{\em}

\newenvironment{prf}{\noindent\textit{Proof:}\begin{mdseries}}{\end{mdseries}{\hfill\scriptsize$\Box$}}

\DeclareMathOperator*{\argmax}{arg\,max}

\floatname{algorithm}{Procedure}

\algrenewcommand\algorithmicindent{1.25em}

\newtheorem{thm}{Theorem}
\newtheorem{lem}{Lemma}
\newtheorem{dfn}{Definition}

\newcommand{\G}{{\mathcal G}}

\DeclareMathOperator{\dis}{dis}

\DeclareMathOperator{\pa}{pa}
\DeclareMathOperator{\de}{de}

\DeclareMathOperator{\ch}{ch}
\DeclareMathOperator{\an}{an}

\DeclareMathOperator{\ant}{ant}
\DeclareMathOperator{\ext}{ext}

\DeclareMathOperator{\nd}{nd}

\DeclareMathOperator{\nb}{nb}

\begin{document}

\twocolumn[

\aistatstitle{General Identification of Dynamic Treatment Regimes Under Interference}

\aistatsauthor{Eli S. Sherman \And David Arbour \And Ilya Shpitser}

\aistatsaddress{Johns Hopkins University \And Adobe Inc. \And Johns Hopkins University} ]

	\begin{abstract}
		In many applied fields, researchers are often interested in tailoring treatments to unit-level characteristics in order to optimize an outcome of interest. Methods for identifying and estimating treatment policies are the subject of the dynamic treatment regime literature. Separately, in many settings the assumption that data are independent and identically distributed does not hold due to inter-subject dependence. The phenomenon where a subject's outcome is dependent on his neighbor's exposure is known as interference. These areas intersect in myriad real-world settings. In this paper we consider the problem of identifying optimal treatment policies in the presence of interference. Using a general representation of interference, via Lauritzen-Wermuth-Freydenburg chain graphs \citep{lauritzen2002chain}, we formalize a variety of policy interventions under interference and extend existing identification theory \citep{tian2008identifying, sherman2018identification}. Finally, we illustrate the efficacy of policy maximization under interference in a simulation study.
	\end{abstract}

	\section{Introduction}
	\label{sec:intro}
	In areas such as precision medicine, economics, and political science, identifying interventions that are optimally tailored to each subject is often of interest. Dynamic treatment regimes (DTRs), which are counterfactual policies used for treatment assignment, represent a promising approach to tailoring treatments. Typically, a causal model is assumed to be known, with analyst-specified exposure and outcome variables. The analyst considers setting the exposure variable according to a treatment policy which is a function of other model variables. She then estimates the counterfactual effect of several candidate policies and picks the one with the best expected outcome. This setup has been extended to sequential settings \citep{laber2014dynamic, chakraborty2013statistical, nabi2018estimation}.

	A key obstacle to obtaining optimal strategies from observational data is identification. The effect of an intervention in a causal model is said to be \emph{identified} if the effect can be expressed as a function of observed data. In algebraic formulations, identification requires carefully enumerating necessary assumptions. In contrast, causal graphical models provide a concise framework for representing assumptions, with numerous general results characterizing identification criteria. In the context of DTRs, \citet{robins1986new} gave an approach for identification of node (i.e., fixed value) and policy interventions in fully-observed directed acyclic graphs (DAGs). \citet{tian2008identifying} and \citet{shpitser2018identification} extended this approach to latent-variable DAG models.

	Orthogonal to treatment customization, classical causal inference assumes independence among study subjects. In many settings, however, subjects' exposures causally affect their neighbors' outcomes. This phenomenon, known as interference \citep{cox1958planning}, has recently attracted substantial attention. \citet{hudgens2008toward} serves as a seminal paper; it defines network-level effects and provides elementary identification conditions. \citet{ogburn2014causal} formalizes DAG representations of interference. Several papers propose relational \citep{maier2013reasoning} or chain graph representations of interference \citep{pena2018unifying, tchetgen2017auto, ogburn2018causal}. \citet{sherman2018identification} is closest to the present work; it explores non-parametric identification in the presence of unobserved confounding.

	A recent paper also considers policies under interference \citep{viviano2019policy}. Our work differs substantially: Viviano focuses on welfare maximization and assumes units are identically distributed. Our characterization of policy interventions generalizes welfare maximization and our network representation is non-parametric.

	\paragraph{Motivating Policies in Networks.}
	In this work, we consider identification of DTRs in the interference setting. As motivation, consider the following example from psephology (the study of elections) \citep{blackwell2013framework}: candidates running for public office target voters by purchasing television advertisements; each candidate must decide how many ads to buy and whether they should be positive (``my record is stellar") or negative (``my opponent is scandalous").
	\begin{figure}[h]
		\begin{center}
			\begin{tikzpicture}[>=stealth, node distance=1.25cm]
			\tikzstyle{format} = [draw, very thick, circle, minimum size=8mm,
			inner sep=0pt]
			\tikzstyle{square} = [draw, very thick, rectangle, minimum size=3.8mm]
			\tikzstyle{unode} = [gray, very thick, circle, minimum size=1.0mm,
			inner sep=0pt]
			
			\begin{scope}
			\path[->, very thick]
			node[] (a) {$A_l$}
			node[above of=a] (c) {$C_l$}
			node[unode, left of=a, yshift=.65cm, xshift=.5cm] (h) {$H_l$}
			
			node[right of=a] (a') {$A_r$}
			node[above of=a'] (c') {$C_r$}
			
			node[below of=a] (y) {$Y_l$}
			node[right of=y] (y') {$Y_r$}
			node[unode, right of=a', yshift=.65cm, xshift=-.5cm] (h') {$H_r$}
			
			(c) edge[blue] (a)
						(c) edge[blue] (a')
			
			(c') edge[blue] (a')
						(c') edge[blue] (a)

			(a) edge[blue] (y)
			(a) edge[blue] (y')
			
			(a') edge[blue] (y)
			(a') edge[blue] (y')
			
			(c) edge[blue, bend right=25] (y)
						(c) edge[blue] (y')
			
						(c') edge[blue] (y)
			(c') edge[blue, bend left=25] (y')
			
			(h) edge[red] (c)
			(h) edge[red] (a)
			(h') edge[red] (a')
			(h') edge[red] (c')

						(y) edge[-, brown] (y')

			node[below of=y, yshift=.75cm, xshift=.625cm] (l) {$(a)$}
			;
			\end{scope}
			
			\begin{scope}[xshift=3cm]
			\path[->, very thick]
			node[] (a) {$A_l$}
			node[above of=a] (c) {$C_l$}
			
			node[right of=a] (a') {$A_r$}
			node[above of=a'] (c') {$C_r$}
			
			node[below of=a] (y) {$Y_l$}
			node[right of=y] (y') {$Y_r$}
			
			(c) edge[blue] (a)
			(c) edge[blue] (a')
			
			(c') edge[blue] (a')
			(c') edge[blue] (a)

			(a) edge[blue] (y)
			(a) edge[blue] (y')
			
			(a') edge[blue] (y)
			(a') edge[blue] (y')
			
			(c) edge[blue] (y')
			(c') edge[blue] (y)
			
			(c) edge[blue, bend right=25] (y)
			
			(c') edge[blue, bend left=25] (y')
			
			(c) edge[<->, red, bend left=25] (a)
			(c') edge[<->, red, bend right=25] (a')

			(y) edge[-, brown] (y')

			node[below of=y, yshift=.75cm, xshift=0.6255cm] (l) {$(b)$}
			;
			\end{scope}
			
			\begin{scope}[xshift=5.5cm]
			\path[->, very thick]
			node[] (a) {$A_l$}
			node[above of=a] (c) {$C_l$}
			
			node[right of=a] (a') {$A_r$}
			node[above of=a'] (c') {$C_r$}
			
			node[below of=a] (y) {$Y_l$}
			node[right of=y] (y') {$Y_r$}
			
			(c) edge[blue] (a)
			(c) edge[blue] (a')
			
			(c') edge[blue] (a')
			(c') edge[blue] (a)

			(a) edge[blue] (y)
			(a) edge[blue] (y')
			
			(a') edge[blue] (y)
			(a') edge[blue] (y')
			
			(c) edge[blue, bend right=25] (y)
			(c) edge[blue] (y')
			
			(c') edge[blue] (y)
			(c') edge[blue, bend left=25] (y')
			

			(a) edge[-, brown] (a')

			node[below of=y, yshift=.75cm, xshift=.625cm] (l) {$(c)$}
			;
			\end{scope}
			
			\end{tikzpicture}
		\end{center}
		\caption{Graphical representations of competitive dynamics in an election campaign, where (a) $H$'s represent latent confounders, (b) is a latent projection with $H$'s replaced by bi-directed edges, and (c) is an alternative model where $A$'s exhibit best-response dynamics.}
		\label{fig:single-shot}
	\end{figure}
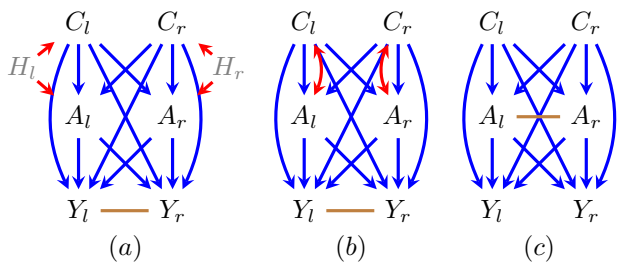

	These dynamics can be represented via the causal graphs in Fig. \ref{fig:single-shot}. For each candidate, $C$ denotes observed pre-decision covariates, such as prior polling performance, previous advertising, and cash on hand,	$A$ represents the candidate's advertising decision, $Y$ represents polling performance in the current decision time frame, and $H$ represents unobserved confounders that affect the candidate's pre-decision covariates and decision but don't directly affect the outcome. $l$ and $r$ index the variables for a left- and right-leaning candidate respectively. Directed edges denote a direct causal relationship, while undirected edges denote non-causal dependence (e.g. $A_l - A_r$ could be interpreted as candidates acting based on beliefs about what each other will do). While we use this two-candidate example as motivation throughout this manuscript, our contributions apply to networks of arbitrary size and topology.

	The remainder of this paper is organized as follows: we fix notation and discuss relevant background work in Secs. \ref{sec:notation} and \ref{sec:background}. We characterize the variety of possible policy interventions in Sec. \ref{sec:varieties}. We then give a novel identification result for effects of policy interventions in Lauritzen-Wermuth-Freydenburg (LWF) latent-variable chain graphs \citep{lauritzen1996graphical, lauritzen2002chain} in Sec. \ref{sec:id}. We demonstrate estimation of these effects via a simulation study in Sec. \ref{sec:experiments} and conclude with a discussion of ongoing work.

\section{Notation}
	\label{sec:notation}
	We first fix notation before describing the task.
	We employ segregated graphs (SGs) \citep{shpitser2015segregated} to represent causal network dynamics.
	SGs are a class of mixed graphical model which are a super-model of latent-variable LWF chain graphs (CGs), which are themselves a super-model of Markov random fields (MRFs) and DAGs. SGs permit three edge types --  undirected ($-$), directed ($\rightarrow$), and bi-directed ($\leftrightarrow$) -- and have the property that no variable has both an incident undirected and bi-directed edge.

	We adopt standard graphical model notation. We denote random variables (interchangeably, vertices in graphs) by capital letters $V$ and their realizations in lowercase $v$, with sets in boldface, $\mathbf{V}$ and $\mathbf{v}$. We use standard genealogical notions for graphical relationships. For a variable $V \in \mathbf{V}$ in a graph $\G$,  parents $\pa_{\G}(V) \equiv \{W \in \mathbf{V} : W \rightarrow V \text{ in } \G\}$, children $\ch_{\G}(V) \equiv \{W \in \mathbf{V} : V \rightarrow W \text{ in } \G\}$, ancestors $\an_{\G}(V) \equiv \{W \in \mathbf{V} : W \rightarrow \dots \rightarrow V \text{ in } \G\}$, descendants $\de_{\G}(V) \equiv \{W \in \mathbf{V} : V \rightarrow \dots \rightarrow W \text{ in } \G\}$, neighbor $\nb_{\G}(V) \equiv \{W \in \mathbf{V} : W - V \text{ in } \G\}$, non-descendant $\nd_{\G}(V) \equiv \mathbf{V} \setminus \de_{\G}(V)$, and district $\dis_{\G}(V) \equiv \{W \in \mathbf{V} : W \leftrightarrow \dots \leftrightarrow V \text{ in } \G\}$.

	Further, the anterior $\ant_{\G}(V)$ is the set of nodes with a partially directed path -- a path containing only $\rightarrow$ and $-$ edges such that no set of undirected edges can be oriented to form a directed cycle -- \emph{into} $V$. The exterior $\ext_{\G}(V)$ is the set of nodes with a partially directed path \emph{out of} $V$. In turn, the
	strict exterior $\overline{\ext}_{\G}(V) \subseteq \ext_{\G}(V)$ omits $V$ and the set $\{W \in \mathbf{V} : W - \dots - V\}$. By convention, $\ext_{\G}(V) \cap \ant_{\G}(V) \cap \dis_{\G}(V) = \{V\}$. These and the above notions can be extended to sets, e.g., for $\mathbf{S} \subseteq \mathbf{V}$, we have $\pa_{\G}(\mathbf{S}) = \cup_{S \in \mathbf{S}} \pa_{\G}(S)$ and, disjunctively, $\pa^s_{\G}(\mathbf{S}) = \pa_{\G}(\mathbf{S}) \setminus \mathbf{S}$. When the relevant graph is clear from context, we drop the $\G$ subscript.
	
	For graphs with a partial ordering $\prec$ on $\mathbf{V}$, let $\mathbf{V}_{\prec A}$ denote $A$'s predecessors in the ordering. For a set $\mathbf{S} \subseteq \mathbf{V}$ in $\G$, let $\G_{\mathbf{S}}$ refers to the subgraph of $\G$ containing only $\mathbf{S}$ and edges connecting nodes in $\mathbf{S}$.

	Finally, we use the notion of a \emph{block} to refer to a set of variables connected by an undirected path. A node with no incident bi-directed nor undirected edges is a \emph{trivial} block \emph{and} a trivial district. The sets of blocks, non-trivial blocks, districts, and cliques in $\G$ are denoted by $\mathcal{B}(\G)$, $\mathcal{B}^{nt}(\G)$, $\mathcal{D}(\G)$, and $\mathcal{C}(\G)$ respectively. In segregated graphs $\mathcal{D}(\G)$ and $\mathcal{B}^{nt}(\G)$ partition $\mathbf{V}$.

	\subsection{Statistical Graphical Models}
	Segregated graphs and their submodels were originally conceived as \emph{statistical} models over random variables, encoding conditional independences in their factorization.
	For instance, a distribution $p(\mathbf{V})$ is `Markov relative to' a CG $\mathcal{G}(\mathbf{V})$ if it factorizes according to the two-level factorization $p(\mathbf{V}) =$
	\begin{align}
	\label{eq:cg-fact}
	\prod_{\mathbf{B} \in \mathcal{B}(\mathcal{G})} p(\mathbf{B} | \pa(\mathbf{B})) = \prod_{\mathbf{B} \in \mathcal{B}(\G)} \frac{ \prod_{\mathbf{C} \in \mathcal{C}^{\star}} \phi_{\mathbf{C}}(\mathbf{C})}{Z(\pa(\mathbf{B}))},
	\end{align}
	where $\mathcal{C}^{\star}
	 = \{\mathbf{C} \in \mathcal{C}((\mathcal{G}^a_{\mathbf{B} \cup \pa_{\G}(\mathbf{B}) }) : \mathbf{C} \not \subseteq \pa_{\mathcal{G}}(\mathbf{B})\}$, and $Z$ is a normalization function. $\mathcal{G}^a_{\mathbf{B} \cup \pa_{\G}(\mathbf{B})}$ is an augmented graph \citep{lauritzen1996graphical}: it is undirected and contains edges in $\G$ between $\mathbf{B}$, edges between nodes in $\pa_{\G}(\mathbf{B})$ and their children in $\mathbf{B}$, and edges between parents. For the corresponding factorizations for DAGs and MRFs, please see the supplementary materials.

	\subsection{Causal Graphical Models}
	In contrast to statistical graphs, causal graphs represent distributions over \emph{counterfactual} variables. For $Y \in \mathbf{V}$ and $\mathbf{A} \subseteq \mathbf{V} \setminus Y$, the counterfactual $Y(\mathbf{a})$ denotes $Y$'s value under the hypothetical scenario in which $\mathbf{A}$ is set to $\mathbf{a}$ via a \emph{node} intervention \citep{pearl2000causality}.

	In this paper, we assume Pearl's functional model.
	In DAGs, counterfactuals $V(\mathbf{a})$ are determined by structural equations $f_{V}(\mathbf{a}, \epsilon_V)$, which remain invariant under an intervention $\mathbf{a}$; $\epsilon_V$ denotes an exogenous random variable for $f_V$. By \emph{recursive substitution}, we can define all other variables in the model: for $\mathbf{A} \subseteq \mathbf{V} \setminus \{V\}$ and $\mathbf{a}$ in the state space of $\mathbf{A}$, $
	p(V(\mathbf{a}))$ (sometimes written as $p(\mathbf{V} | \text{do}(\mathbf{a})$ \citep{pearl2000causality}) is defined as $V(\mathbf{a}_{\pa(V)}, \{W(\mathbf{a}) : W \in \pa(V) \setminus \mathbf{A}\})$.
	
	Causal CGs follow similar semantics. Each variable $B$ in a block $\mathbf{B}$ is determined by a structural equation $f_B(\mathbf{B} \setminus \{B\}, \pa(\mathbf{B}), \epsilon_{B})$, a function of other variables in $\mathbf{B}$, the parents of $\mathbf{B}$, and an exogenous variable. Each $\mathbf{B}$'s joint distribution $\mathbf{B}$ is obtained by Gibbs sampling over the structural equations for $\mathbf{B}$ until equilibrium (trivial blocks equilibrate instantly). Assuming an ordering on blocks in $\G$, but not on variables in each block, and iid realizations of $\epsilon_{B_i}$, the data generating process for CGs is given by Procedure \ref{proc:gibbs} \citep{lauritzen2002chain}.
	\begin{algorithm}[h]
		\caption{CG Data Generating Process}\label{alg:gfa}
		\begin{algorithmic}[1]
			\Procedure{CG-DGP}{$\G, \{f_B : B \in \mathbf{V}\}$}
			\ForAll{block $\mathbf{B}_i \in \mathcal{B}(\G)$}
				\Repeat{
				\ForAll{variable $B_j \in \mathbf{B}_i$}
					\State{$B_j \leftarrow f_{B_j}(\mathbf{B}_i \setminus B_j, \pa_{\G}(\mathbf{B}_i), \epsilon_{B_j})$}
				\EndFor}
				\Until{equilibrium}
			\EndFor
			\Return $\mathbf{V}$
			\EndProcedure
		\end{algorithmic}
		\label{proc:gibbs}
	\end{algorithm}
	
	A parameter is \emph{identifiable} in a causal model if it can be expressed as a function of observed data. In fully observed DAGs and CGs, all node intervention counterfactuals are identified by the \emph{g-formula} \citep{robins1986new} and \emph{chain graph g-formula} \citep{lauritzen2002chain} respectively (first two rows of Table \ref{tab:id-simplification}).

\section{Identification in Latent-Variable Causal Graphical Models}
\label{sec:background}
	In this section, we review identification theory in latent variable causal models. The current work bridges these literatures: we posit a sound and complete algorithm for the identification of responses to policies in latent variable (LV) causal CGs.
\begin{table*}[t]
	\begin{center}
		\begin{tabular}{|l|l|l|l|p{6.2cm}|}
			\hline
			Graph Type & Latents & Intervention Type & $\mathbf{Y}^{\star}$ & Modified Factorization\\ \hline \hline
			DAG & No & Node -- $\mathbf{a}$ & N/A & $\prod_{V \in \mathbf{V} \setminus \mathbf{A}} p(V | \pa(V))|_{\mathbf{A} = \mathbf{a}}$\\ \hline
			CG & No & Node -- $\mathbf{a}$ & N/A & $\prod_{\mathbf{B} \in \mathcal{B}(\G)} p(\mathbf{B} \setminus \mathbf{A} | \pa(\mathbf{B}), \mathbf{B} \cap \mathbf{A})|_{\mathbf{A} = \mathbf{a}}$\\ \hline
			ADMG & Yes & Node -- $\mathbf{a}$ & $\an_{\G_{\mathbf{V} \setminus \mathbf{A}}}(\mathbf{Y})$ & $\prod_{\mathbf{D} \in \mathcal{D}(\G_{\mathbf{Y}^{\star}})} \phi_{\mathbf{V} \setminus \mathbf{D}}(p(\mathbf{V}); \G)|_{\mathbf{A} = \mathbf{a}}$\\ \hline
			SG & Yes & Node -- $\mathbf{a}$ & $\ant_{\G_{\mathbf{V} \setminus \mathbf{A}}}(\mathbf{Y})$ & $\prod_{\mathbf{D} \in \mathcal{D}(\tilde{\G}^d)} \phi_{\mathbf{D}^{\star} \setminus \mathbf{D}} (q(\mathbf{D}^{\star} | \pa^s_{\G}(\mathbf{D}^{\star})); \G^d)
			\times \prod_{\mathbf{B} \in \mathcal{B}(\tilde{\G}^b)} p(\mathbf{B} \setminus \mathbf{A} | \pa_{\G_{\mathbf{Y}^{\star}}}(\mathbf{B}), \mathbf{B} \cap \mathbf{A})|_{\mathbf{A} = \mathbf{a}}$\\ \hline
			ADMG & Yes & Policy -- $\mathbf{f_A}$ & $\an_{\G_{\mathbf{f_A}}}(\mathbf{Y})$ & $\prod_{\mathbf{D} \in \mathcal{D}(\G_{\mathbf{Y}^{\star}})} \phi_{\mathbf{V} \setminus \mathbf{D}}(p(\mathbf{V}); \G)|_{\mathbf{A} = \mathbf{\tilde{a}}}$\\ \hline
		\end{tabular}
	\caption{Summary of existing identification approaches. The first two rows use standard g-formulas, the third row is the ID algorithm, and the final two extend ID. The present work generalizes the last two rows. In the fifth row, $\mathbf{\tilde{a}} = \{A = f_A(\mathbf{W}_A) | A \in \pa_{\G}(\mathbf{D}) \cap \mathbf{A}\}$ if $\pa_{\G}(\mathbf{D}) \cap \mathbf{A} \neq \emptyset$ and $\tilde{\mathbf{a}} = \emptyset$ otherwise.}
	\label{tab:id-simplification}
	\end{center}
\end{table*}

	\subsection{Re-expressing the ID Algorithm}
	\citet{tian2002general} gave a general condition for identification of node interventions in latent-variable DAGs. \citet{shpitser2006identification} re-expressed Tian's condition as a concise algorithm and proved that it is complete. Recently, \citet{richardson2017nested} rephrased the algorithm in terms of a recursive fixing operator which acts as a modified nested Markov factorization.

	\citet{richardson2017nested} makes clear the connections between the ID algorithm, which is a modified nested factorization of acyclic directed mixed graphs (ADMGs), and the g-formula (Table \ref{tab:id-simplification}, first row), which is a modified DAG factorization. This formalism enables straightforward generalizations to other identification settings. For these reasons, we base our SG policy identification results on this framework. The framework relies on several concepts which we highlight here; each existing ID approach is summarized in Table \ref{tab:id-simplification}. For a complete treatment, please see the supplement.

	\paragraph{Latent Projections.}
	Rather than considering LV-DAGs explicitly, \citet{richardson2017nested} considers ADMGs. ADMGs permit directed and bi-directed edges and represent equivalence classes of LV-DAGs. Given an LV-DAG $\G(\mathbf{V} \cup \mathbf{H})$, with $\mathbf{V}$ observed and $\mathbf{H}$ latent, the corresponding ADMG $\G(\mathbf{V})$ is obtained via a latent projection operation \citep{verma1991equivalence}. For example, Fig. \ref{fig:single-shot}(b) is the latent projection of Fig. \ref{fig:single-shot}(a). We also define conditional ADMGs (CADMGs), which partition nodes into random $\mathbf{V}$ and fixed $\mathbf{W}$ variables. CADMGs with $\mathbf{W} = \emptyset$ are trivially ADMGs.

	Segregated graphs are the chain graph analogue of ADMGs, where SGs represent an equivalence class of LV-CGs. For a latent variable CG $\G(\mathbf{V} \cup \mathbf{H})$, $\mathbf{H}$ is \emph{block-safe} \citep{sherman2018identification} if no $V \in \mathbf{V}$ has a latent parent and no latent $H \in \mathbf{H}$ has an incident undirected edge. By applying the same latent projection operation mentioned above to a LV-CG with block-safe $\mathbf{H}$, one obtains the corresponding SG.

	\paragraph{Kernels and Fixing.}
	Whereas DAGs and CGs factorize as products of conditional distributions, ADMGs and SGs factorize as products of \emph{kernels} \citep{lauritzen1996graphical}. A kernel $q_{\mathbf{V}}(\mathbf{V} | \mathbf{W})$ is a function, mapping values of $\mathbf{W}$ to normalized densities on $\mathbf{V}$. For some $\mathbf{A} \subseteq \mathbf{V}$, conditioning and marginalization are defined as:
	\begin{align*}
		q(\mathbf{A} | \mathbf{W}) \equiv \sum_{\mathbf{V} \setminus \mathbf{A}} q(\mathbf{V} | \mathbf{W}); \hspace{.2cm} q(\mathbf{V} \setminus \mathbf{A} | \mathbf{A}, \mathbf{W}) \equiv \frac{q(\mathbf{V} | \mathbf{W})}{q(\mathbf{A} | \mathbf{W})}.
	\end{align*}
	The notion of fixing variables is closely tied to kernels. In a CADMG $\G(\mathbf{V},\mathbf{W})$, a variable $V$ is fixable if $\de(V) \cap \dis(V) = \emptyset$. In a DAG $\G$ with corresponding distribution $p(\mathbf{V})$, fixing $V$ corresponds to applying the g-formula to obtain a new distribution $p(\mathbf{V} \setminus V)$ and a new graph $\G'$. For a CADMG $\G(\mathbf{V}, \mathbf{W})$ with corresponding kernel $q(\mathbf{V} | \mathbf{W})$, \citet{richardson2017nested} defines similar operations, denoted $\phi_V(\G)$ and $\phi_V(q; \G)$. These operators yields a new CADMG $\G'(\mathbf{V} \setminus \{V\}, \mathbf{W} \cup \{V\})$ in which all
	edges \emph{into} $V$ are removed and a new kernel
		$q'(\mathbf{V} \setminus \{V\} | \mathbf{W} \cup \{V\}) \equiv \frac{q(\mathbf{V} | \mathbf{W})}{q(\mathbf{V} | \pa_{\G}(\dis_{\G}(V)) \cup \dis_{\G}(V), \mathbf{W})}$, respectively. These operators were used to define the nested Markov model.


	Fixability also extends to sets of variables $\mathbf{S} \subseteq \mathbf{V}$ in a ADMG $\G(\mathbf{V})$ when $q_{\mathbf{V}}$ is in the nested Markov model. If it is possible to find a sequence $S_1, S_2, \dots$ of the variables in $\mathbf{S}$ such that $S_1$ is fixable in $\G$, $S_2$ is fixable in $\phi_{S_1}(q_{\mathbf{V}}; \G)$ and so on, then $\mathbf{S}$ is fixable and $\mathbf{V} \setminus \mathbf{S}$ is said to be \emph{reachable} in $\G$.
	Since all valid fixing sequences on $\mathbf{S}$ yield the same CADMG $\G(\mathbf{V} \setminus \mathbf{S}, \mathbf{S})$ via $\phi$, and (if $p(\mathbf{V})$ is nested Markov with respect to $\G(\mathbf{V})$), all fixing sequences on $\mathbf{S}$ valid in $\G(\mathbf{V})$ yield the same kernel $q(\mathbf{V} \setminus \mathbf{S} | \mathbf{S})$ via $\phi$, the fixing operators can be defined for sets unambiguously: $\phi_{\mathbf{S}}(\G)$ and $\phi_{\mathbf{S}}(q; \G)$.


	This notation permits reformulating the ID algorithm. For an ADMG $\G(\mathbf{V})$, let $\mathbf{Y}, \mathbf{A} \subseteq \mathbf{V}$ be disjoint and $\mathbf{Y}^{\star} \equiv \an_{\G_{\mathbf{V} \setminus \mathbf{A}}}(\mathbf{Y})$. $p(\mathbf{Y} | \text{do}(\mathbf{a}))$ is identified in $\G$ if and only if every district $\mathbf{D} \in \mathcal{D}(\G_{\mathbf{Y}^{\star}})$ is reachable in $\G$. If identified, $p(\mathbf{Y} | \text{do}(\mathbf{a}))$ is given by summing the modified factorization in row three of Table \ref{tab:id-simplification} over $\mathbf{Y}^{\star} \setminus \mathbf{Y}$. 
	
	Returning to our elections example (Fig. \ref{fig:single-shot}(c)), suppose we assume each candidate's decision is independent of other decisions given covariates (i.e., no $A_l - A_r$ edge). We can use this formula to consider the effect on a candidate's polling of advertising positively and negatively in fixed proportion (say, equally, $a = .5$).
	
	As another example, consider the subgraph on $C_1, A_1, M_1, Y_1$ in Fig. \ref{fig:gross-sg}(a); $p(Y_1|\text{do}(a_1))$ is not identified \citep{shpitser2006identification}. In the $C_2, A_2, M_2, Y_2$ subgraph, however, $p(Y_2 | \text{do}(a_2))$ is identified by the front-door formula:
	\begin{align*}
		\sum_{M_2, C_2} p(M_2 | a_2, C_2) p(C_2) \sum_{A_2'} p(Y_2 | M_2, C_2, A_2') p(A_2' | C_2)
	\end{align*}

	\subsection{Identification in Segregated Graphs}
	\paragraph{The Segregated Factorization.}
	Extending the factorizations for ADMGs and CGs, \citet{sherman2018identification} defines the segregated factorization for SGs.
	
	Recall that an SG $\G$ is partitioned by variables that lie in non-trivial blocks, denoted $\mathbf{B}^{\star} = \cup_{\mathbf{B} \in \mathcal{B}^{nt}(\G)} \mathbf{B}$, and those that don't, denoted $\mathbf{D}^{\star} = \cup_{\mathbf{D} \in \mathcal{D}(\G)} \mathbf{D}$.
	An SG satisfying the segregated factorization can be expressed as the product of kernels for these two sets. 

	The first kernel, $q(\mathbf{B}^{\star} | \pa^s_{\G}(\mathbf{B}^{\star})) = \prod_{\mathbf{B} \in \mathcal{B}^{nt}(\G)} p(\mathbf{B} | \pa_{\G}(\mathbf{B}))$, factorizes with respect to a conditional chain graph (CCG) $\G(\mathbf{V}, \mathbf{W})$, 
	which we denote by $\G^b$ with $\mathbf{V}$ corresponding to $\mathbf{B}^{\star}$ and $\mathbf{W}$ to $\pa^s(\mathbf{B}^{\star})$. $\G^b$ contains edges between nodes in $\mathbf{B}^{\star}$ and between nodes in $\pa^s_{\G}(\mathbf{B}^{\star})$ that exist in $\G$.

	The second kernel, $q(\mathbf{D}^{\star} | \pa^s_{\G}(\mathbf{D}^{\star})) = \frac{p(\mathbf{V})}{q(\mathbf{B}^{\star} | \pa^s_{\G}(\mathbf{B}^{\star}))}$, nested factorizes with respect to a CADMG denoted $\G^d$, with random nodes $\mathbf{D}^{\star}$ and fixed nodes $\pa^s(\mathbf{D}^{\star})$. Like $\G^b$, $\G^d$ contains edges between nodes in $\mathbf{D}^{\star}$ and between nodes in $\pa^s_{\G}(\mathbf{D}^{\star})$ that are present in $\G$.
	
\begin{figure*}[t!]
	\begin{center}
		\begin{tikzpicture}[>=stealth, node distance=1.01cm]
		\tikzstyle{format} = [draw, very thick, circle, minimum size=8mm,
		inner sep=0pt]
		\tikzstyle{square} = [draw, very thick, rectangle, minimum size=3.8mm]
		\tikzstyle{unode} = [draw, gray, very thick, circle, minimum size=1.0mm,
		inner sep=0pt]
		
		\begin{scope}
		\path[->, very thick]
		node[] (a1) {$A_1$}
		node[below of=a1] (m1) {$M_1$}
		node[above of=a1] (c1) {$C_1$}
		node[below of=m1] (y1) {$Y_1$}
		
		node[right of=a1] (a2) {$A_2$}
		node[below of=a2] (m2) {$M_2$}
		node[above of=a2] (c2) {$C_2$}
		node[below of=m2] (y2) {$Y_2$}
		
		node[right of=a2] (a3) {$A_3$}
		node[below of=a3] (m3) {$M_3$}
		node[above of=a3] (c3) {$C_3$}
		node[below of=m3] (y3) {$Y_3$}

		(c1) edge[blue] (a1)
		(a1) edge[blue] (m1)
		(m1) edge[blue] (y1)
		
		(c2) edge[blue] (a2)
		(a2) edge[blue] (m2)
		(m2) edge[blue] (y2)
		(c2) edge[blue, bend right=20] (m2)
		(c2) edge[blue, bend left=25] (y2)
		
		(c3) edge[blue] (a3)
		(a3) edge[blue] (m3)
		(m3) edge[blue] (y3)
		
		(y2) edge[<->, red] (y3)
		
		(a1) edge[blue, bend right=20] (y1)
		(a1) edge[<->, red, bend right=40] (y1)
		
		(a2) edge[<->, red, bend left=40] (y2)
		
		(c1) edge[<->, red, bend left=25] (y1)
		
		(m1) edge[-, brown] (m2)
		(m2) edge[-, brown] (m3)
		
		(c2) edge[-, brown] (c3)
		(c2) edge[blue] (m3)
		
		node[below of=y2, yshift=.4cm] (l) {$(a)\,\, \G$}
		;
		\end{scope}
		
		\begin{scope}[xshift=3.5cm]
		\path[->, very thick]
		node[square] (a1) {$A_1$}
		node[below of=a1] (m1) {$M_1$}
		node[square, above of=a1] (c1) {$C_1$}
		node[square, below of=m1] (y1) {$Y_1$}
		
		node[square, right of=a1] (a2) {$A_2$}
		node[below of=a2] (m2) {$M_2$}
		node[above of=a2] (c2) {$C_2$}
		node[square, below of=m2] (y2) {$Y_2$}
		
		node[square, right of=a2] (a3) {$A_3$}
		node[below of=a3] (m3) {$M_3$}
		node[above of=a3] (c3) {$C_3$}
		node[square, below of=m3] (y3) {$Y_3$}

		(a1) edge[blue] (m1)
		
		(a2) edge[blue] (m2)
		(c2) edge[blue, bend right=35] (m2)
		
		(a3) edge[blue] (m3)
		
		
		
		
		
		(m1) edge[-, brown] (m2)
		(m2) edge[-, brown] (m3)
		
		(c2) edge[-, brown] (c3)
		(c2) edge[blue] (m3)
		
		node[below of=y2, yshift=.4cm] (l) {$(b)\,\, \G^b$}
		;
		\end{scope}
		
		\begin{scope}[xshift=7cm]
		\path[->, very thick]
		node[] (a1) {$A_1$}
		node[square, below of=a1] (m1) {$M_1$}
		node[above of=a1] (c1) {$C_1$}
		node[below of=m1] (y1) {$Y_1$}
		
		node[right of=a1] (a2) {$A_2$}
		node[square, below of=a2] (m2) {$M_2$}
		node[square, above of=a2] (c2) {$C_2$}
		node[below of=m2] (y2) {$Y_2$}
		
		node[right of=a2] (a3) {$A_3$}
		node[square, below of=a3] (m3) {$M_3$}
		node[square, above of=a3] (c3) {$C_3$}
		node[below of=m3] (y3) {$Y_3$}

		(c1) edge[blue] (a1)
		(m1) edge[blue] (y1)
		
		(c2) edge[blue] (a2)
		(m2) edge[blue] (y2)
		(c2) edge[blue, bend left=30] (y2)
		
		(c3) edge[blue] (a3)
		(m3) edge[blue] (y3)
		
		(y2) edge[<->, red] (y3)
		
		(a1) edge[blue, bend right=35] (y1)
		(a1) edge[<->, red, bend right=45] (y1)
		
		(a2) edge[<->, red, bend left=40] (y2)
		
		(c1) edge[<->, red, bend left=25] (y1)
		
		
		
		node[below of=y2, yshift=.4cm] (l) {$(c)\,\, \G^d$}
		;
		\end{scope}
		
		\begin{scope}[xshift=10.5cm]
		\path[->, very thick]
		node[] (a1) {$A_1$}
		node[below of=a1] (m1) {$M_1$}
		node[above of=a1] (c1) {$C_1$}
		node[below of=m1] (y1) {$Y_1$}
		
		node[right of=a1] (a2) {$A_2$}
		node[below of=a2] (m2) {$M_2$}
		node[above of=a2] (c2) {$C_2$}
		node[below of=m2] (y2) {$Y_2$}
		
		node[right of=a2] (a3) {$A_3$}
		node[below of=a3] (m3) {$M_3$}
		node[above of=a3] (c3) {$C_3$}
		node[below of=m3] (y3) {$Y_3$}

		(c1) edge[dashed, blue] (a1)
		(a1) edge[blue] (m1)
		(m1) edge[blue] (y1)
		
		(c2) edge[blue, dashed] (a2)
		(a2) edge[blue, dashed] (m2)
		(m2) edge[blue] (y2)
		(c2) edge[blue, dashed, bend right=20] (m2)
		(c2) edge[blue, bend left=25] (y2)
		
		(c3) edge[blue, dashed] (a3)
		(a3) edge[blue] (m3)
		(m3) edge[blue] (y3)
		
		(y2) edge[<->, red] (y3)
		
		(a1) edge[blue, bend right=20] (y1)
		
		
		(m1) edge[<-, dashed, blue] (m2)
		(m2) edge[-, dashed, brown] (m3)
		
		(c2) edge[dashed, blue] (a1)
		(c3) edge[dashed, blue] (a2)
		
		(a2) edge[-, dashed, brown] (a3)
		
		(c2) edge[-, brown] (c3)
		(c2) edge[blue] (m3)
		
		node[below of=y2, yshift=.4cm] (l) {$(d)\,\, \G_{\mathbf{f_A}}$}
		;
		\end{scope}
		
		\begin{scope}[xshift=13cm]
		\path[->, very thick]
		node[] (a1) {}
		node[below of=a1] (m1) {}
		node[above of=a1] (c1) {}
		node[below of=m1] (y1) {}
		
		node[right of=a1] (a2) {}
		node[below of=a2] (m2) {}
		node[above of=a2] (c2) {$C_2$}
		node[below of=m2] (y2) {$Y_2$}
		
		node[right of=a2] (a3) {}
		node[below of=a3] (m3) {$M_3$}
		node[above of=a3] (c3) {$C_3$}
		node[below of=m3] (y3) {$Y_3$}
		
		(y2) edge[<->, red] (y3)
		
		
		(m2) edge[blue] (y2)
		(c2) edge[blue] (y2)
		
		(m3) edge[blue] (y3)
		
		
		
		
		(c3) edge[-, brown] (c2)
		(c2) edge[blue] (m3)
		
		node[below of=y2, yshift=.4cm, xshift=.625cm] (l) {$(e) \,\, \G_{\mathbf{Y}^{\star}}$}
		;
		\end{scope}
		
		\end{tikzpicture}
	\end{center}
	\caption{(\ref{fig:gross-sg}(a) An SG $\G$ where bi-directed edges signify the presence of latent confounders. \ref{fig:gross-sg}(b) and (c) The conditional chain graph $\G^b$ and conditional ADMG $\G^d$ obtained from $\G$. \ref{fig:gross-sg}(d) The post-intervention graph $\G_{\mathbf{f_A}}$ induced by the policy intervention $\mathbf{f_A}$ as described in Sec. \ref{sec:varieties}. Nodes with changed structural equations have dashed incoming edges. \ref{fig:gross-sg}(e) The corresponding $\G_{\mathbf{Y}^{\star}}$ for $\G_{\mathbf{f_A}}$ in \ref{fig:gross-sg}(b) with outcome $\mathbf{Y} = \{Y_2, Y_3\}$.}
	\label{fig:gross-sg}
\end{figure*}
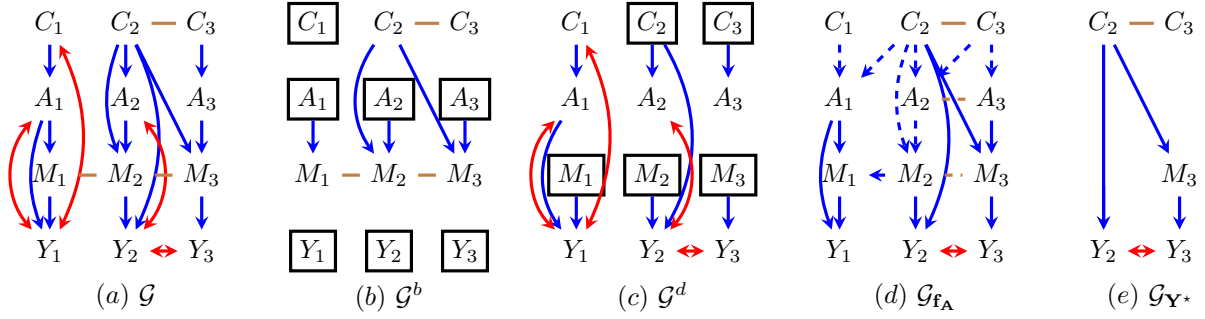

	For example, in the graph in Fig. \ref{fig:gross-sg}(a), we have
	\begin{align*}
		q(\mathbf{D}^{\star} | \pa_{\G}^s(\mathbf{D}^{\star})) &= p(Y_2, Y_3, A_2 | C_2, M_2, M_3)\\
		&\times p(Y_1, A_1, C_1 | M_1) p(A_3 | C_3)
	\end{align*}
	\begin{align*}
		q(\mathbf{B}^{\star} | \pa_{\G}^s(\mathbf{B}^{\star})) &= p(M_1, M_2, M_3 | A_1, A_2, A_3) p(C_2, C_3)
	\end{align*}
	which correspond to Fig. \ref{fig:gross-sg}(b) and (c) respectively.

	\paragraph{The Segregated Graph ID Algorithm.}
	We can now describe an extension of the ID algorithm for node interventions in segregated graphs. For a SG $\G(\mathbf{V})$, fix disjoint $\mathbf{Y}, \mathbf{A} \subseteq \mathbf{V}$. Let $\mathbf{Y}^{\star} \equiv \ant_{\G_{\mathbf{V} \setminus \mathbf{A}}} (\mathbf{Y})$. Define $\tilde{\G}^d$ and $\tilde{\G}^b$ to be the CADMG and CCG respectively obtained from $\G_{\mathbf{Y}^{\star}}$.
	$p(\mathbf{Y} | \text{do}(\mathbf{a}))$ is identified in $\G$ if and only if each $\mathbf{D} \in \mathcal{D}(\tilde{\G}^d)$ is reachable in $\G^d$. If identified, $p(\mathbf{Y} | \text{do}(\mathbf{a}))$ is equal to the modified factorization in row four of Table \ref{tab:id-simplification}, summed over $\mathbf{Y}^{\star} \setminus \mathbf{Y}$.

	Coming back to our elections example, Fig. \ref{fig:single-shot}(b), this formula is applicable when considering the effect of the left-leaning candidate taking a fixed action $a_l$, with the right-leaning candidate's action still having an impact on the left's poll standing. $p(Y_l(a_l))$ is identified by:
	\begin{align*}
		\sum_{C_l, C_r, A_r, Y_r} \!\!\!\!\!p(Y_l, Y_r | C_l, C_r, A_r, a_l) p(A_2 | C_l, C_r) p(C_l) p(C_r)
	\end{align*}

	\subsection{Policy Interventions in ADMGs}
	Extending node interventions, we now consider \emph{policy interventions}. For an ADMG $\mathcal{G}(\mathbf{V})$ with topological ordering $\prec$ on $\mathbf{V}$ and an intervention set $\mathbf{A} \subseteq \mathbf{V}$, let $\mathbf{f}_{\mathbf{A}}$ be the set of policies $\{f_A : A \in \mathbf{A}\}$. Each $f_A$ is a stochastic function of some $\mathbf{W}_A \subseteq \mathbf{V}_{\prec A}$, where $f_A(\mathbf{W}_A)$ maps the state space of $\mathbf{W}_{A}$ to the state space of $A$. Intervening with $f_A$ corresponds to removing edges \emph{into} $A$ in $\mathcal{G}$ and adding edges from $\mathbf{W}_A$ to $A$, yielding a new graph $\mathcal{G}_{\mathbf{f_A}}$.


	\citet{tian2008identifying} gave a policy-analogue of the ID algorithm for $p(\{\mathbf{V} \setminus \mathbf{A}\} (\mathbf{f}_{\mathbf{A}}))$,
	which \citet{shpitser2018identification} re-expressed via the fixing operator $\phi$. Let $\mathbf{Y}^{\star} \equiv \an_{\G_{\mathbf{f_A}}}(\mathbf{Y}) \setminus \mathbf{A}$. A policy-analogue of the ID algorithm follows: $p(\mathbf{Y}(\mathbf{f_A}))$ is identified in $\G$ if and only if $p(\mathbf{Y}^{\star}(\mathbf{a}))$ is identified in $\G$; if identified, $p(\mathbf{Y}(\mathbf{f_A}))$ is obtained by summing over $(\mathbf{Y}^{\star} \cup \mathbf{A}) \setminus \mathbf{Y}$ in the modified factorization in row five of Table \ref{tab:id-simplification}. 
	
	In our elections example, assume candidates' decisions and outcomes are independent of each other. This formula can be used to consider the effect on a candidate's polling of advertising based on the relevant covariates, e.g., if the election is less than 2 months away, advertise negatively, and buy positive ads until then.

\section{Varieties of Policy Interventions}
\label{sec:varieties}
	We now describe extensions of policy interventions to network data representable by SGs. These interventions correspond to replacing structural equations in Procedure \ref{proc:gibbs} with new equations, under conditions we describe below such that the resulting data generating process yields a new SG. As we discuss, these policy interventions induce a variety of edge changes in SGs.

	\subsection{Inducing Direct Causation}
	As in the latent-variable DAG case \citep{shpitser2018identification}, we can intervene by inducing a parent-child relationship between the treatment node and other variables in the graph or modify the nature of existing relationship. In our elections example from Sec. \ref{sec:intro}, this might correspond to intervening on the left candidate's decision $A_l$ such that she adopts a new strategy for responding to her competitor's characteristics $C_r$ relative to her (observed) status quo strategy. For illustrative purposes, this type of intervention is demonstrated by the addition of the $C_2 \rightarrow A_1$ edge and the modification to the $C_1 \rightarrow A_1$ edge between Fig. \ref{fig:gross-sg}(a) and \ref{fig:gross-sg}(d).

	\subsection{Inducing or Modifying Undirected Dependence}
	We can also consider changing the block structure of the SG. There are two types of such interventions:
 	\begin{enumerate}[wide, labelwidth=!, labelindent=0pt]

			\item \textit{Modifying the functional form encoded by an existing undirected edge}. In Fig. \ref{fig:single-shot} (b), we can think of the undirected edge $A_l - A_r$ as representing each candidates' beliefs about the other candidate's actions. In the observed data, candidates will best-respond to each other according to these beliefs. We can imagine changing the way one (or both) of the candidates reasons about their opponent's possible actions, such as making one candidate hyper-responsive to their opponent's anticipated action. Mechanically, we intervene on $A_l$ (analogously $A_r$) with a function $f_{A_l}$ that takes $A_r$ as an argument. We needn't intervene on the other candidate to maintain the undirected edge between the $A$'s. This type of intervention is demonstrated by the change to the $M_2 - M_3$ edge from Fig. \ref{fig:gross-sg}(a) to \ref{fig:gross-sg}(d).

			\item \textit{Inducing co-dependence by adding a new undirected edge between two nodes}. This might correspond to having a third candidate $c$ join the race and intervening such that $A_c - A_l$ and $A_c - A_r$. In this case, it is necessary to intervene on both endpoint nodes for the new undirected edge in order; we modify the respective structural equations to take the other endpoint as an argument. We further restrict these interventions by requiring that they do not induce a partially directed cycle, which would violate the segregation property of the graph. We formalize this requirement below. We note that this type of intervention can be thought of as a chain graph generalization of connection interventions, proposed in \citet{sherman2019intervening}. As an example, consider the addition of the $A_2 - A_3$ edge in Fig. \ref{fig:gross-sg}(d) relative to \ref{fig:gross-sg}(a).
		\end{enumerate}
	
	\subsection{Removing Dependence}
	Finally, we can consider removing undirected dependence between nodes. Once again there are two types:
	\begin{enumerate}[wide, labelwidth=!, labelindent=0pt]
		\item \textit{Partial removal.} We intervene on a single node to make its structural equation no longer a function of the other end point of the undirected edge. In our elections example (Fig. \ref{fig:single-shot}(c)), this corresponds to a `first mover' scenario where $A_l$ is made to not depend on $A_r$ and thus candidate $l$ makes her decision before candidate $r$. Graphically, we change the undirected edge $A_l - A_r$ to a directed edge $A_l \rightarrow A_r$ since $A_r$ is still determined by candidate $l$'s decision; see, for instance, the $M_1 - M_2$ and $M_1 \leftarrow M_2$ edges in Fig. \ref{fig:gross-sg}(a) and \ref{fig:gross-sg}(d).

		\item \textit{Complete removal.} We remove both dependences by intervening on both endpoints of an undirected edge so that the structural equations are no longer functions of each other. This corresponds to a candidate dropping out of the race in our elections example. Like dependence-inducing interventions above, this intervention type can be viewed as an SG analogue of severance interventions \citep{sherman2019intervening}.
	\end{enumerate}

	\section{Identification of Policies in Segregated Graphs}
	\label{sec:id}
	In this section we formalize policy interventions and provide a procedure for obtaining the post-intervention graph from $\G$. We then give a criterion for the identification of policy interventions in SGs \citep{shpitser2015segregated} and demonstrate application of this criterion to Fig. \ref{fig:gross-sg} and to our electoral example, Fig. \ref{fig:single-shot}. We defer proofs and derivations to the supplement.
	
	\begin{algorithm}[t]
		\caption{Obtaining $\G_{\mathbf{f_A}}$ from $\G$}\label{alg:gfa}
		\begin{algorithmic}[1]
			\Procedure{InterveneGraph}{$\G, \mathbf{f_A}(\mathbf{Z_A})$}
			\State{Initialize $\G_{\mathbf{f_A}} \leftarrow \G$}
			\ForAll{$A \in \mathbf{A}$}
			\State{Replace all $V - A$ with $A \rightarrow V$ in $\G_{\mathbf{f_A}}$}
			\State{Remove all $\cdot \rightarrow A$, $\cdot \leftrightarrow A$ from $\G_{\mathbf{f_A}}$}
			\State{Add edges $\mathbf{Z}_A \rightarrow A$ in $\G_{\mathbf{f_A}}$}
			\EndFor
			\ForAll{$V_i, V_j \in \mathbf{V}$}
			\If{$V_i \rightarrow V_j$ and $V_j \rightarrow V_i$ in $\G_{\mathbf{f_A}}$}
			\State{Remove $V_i \rightarrow V_j$ and $V_j \rightarrow V_i$ from $\G_{\mathbf{f_A}}$}
			\State{Add $V_i - V_j$ in $\G_{\mathbf{f_A}}$}
			\EndIf
			\EndFor
			\Return $\G_{\mathbf{f_A}}$
			\EndProcedure
		\end{algorithmic}
		\label{proc:gfa}
	\end{algorithm}

	\subsection{Formalizing Policy Interventions in Segregated Graphs}
	Before providing identification conditions, we first formally define policy interventions in SGs.
	Recall that in ADMGs a policy $f_A(\mathbf{W}_A) \in \mathbf{f}_{\mathbf{A}}$ was required to be a function of variables $\mathbf{W}_A$ preceding $A$ in a topological ordering on the nodes in $\G$. In SGs we loosen this restriction such that $f_A$ operates as a structural equation that can also be a stochastic function of variables in the same block as $A$. For an intervention inducing a block or modifying the structural equations in a block, we use Procedure \ref{proc:gibbs} to obtain a new block distribution.
	
	For $f_A(\mathbf{Z}_A)$ to be a valid policy in an SG $\G(\mathbf{V})$, we require $\mathbf{Z}_A \subseteq \mathbf{V} \setminus \overline{\ext}(A)$. In turn, for $\mathbf{f}_{\mathbf{A}}$ to be valid, all constituent policies must be valid and they may not collectively violate the CG property by inducing a partially directed cycle. We formalize this notion as follows: let $A_i \triangle A_j$ denote that variable $A_i$ is made (either directly or indirectly) a function of $A_j$ for $A_i, A_j \in \mathbf{A}$. To prevent partially directed cycles, we stipulate that if $A_i \triangle A_j$ and $A_j \triangle A_i$ then we require $A_i \in \mathbf{Z}_{A_j}$ and vice versa. This motivates the following definition.

	\begin{dfn}
		A policy intervention $\mathbf{f_A}(\mathbf{Z_A})$ is `segregation preserving' if (a) for each $A \in \mathbf{A}$, $\mathbf{Z}_A \subseteq \mathbf{V} \setminus \overline{\ext}(A)$, and (b) for any $A_i, A_j \in \mathbf{A}$ if $A_i \triangle A_j$ and $A_j \triangle A_i$, we have that $A_i \in \mathbf{Z}_{A_j}$ and $A_j \in \mathbf{Z}_{A_i}$.
	\end{dfn}

	For a given intervention set $\mathbf{f_A}$, we can construct a post-intervention graph $\G_{\mathbf{f_A}}$ according to Procedure \ref{proc:gfa}, which follows from the analogous procedure for policy identification in LV-DAGs. In Lemma \ref{lem:g-fa}, we show that $\G_{\mathbf{f_A}}$ is an SG when $\mathbf{f_A}$ is segregation-preserving. As an example of this procedure's application, consider Fig. \ref{fig:gross-sg}(a). Suppose we wish to perform an intervention $\mathbf{f_A}(\mathbf{Z_A})$ as in Table \ref{tab:intervention}. Then $\G_{\mathbf{f_A}}$ is given by Fig. \ref{fig:gross-sg}.

	\begin{table}[h]
		\begin{center}
			\begin{tabular}{|l||l|l|l|l|}
				\hline
				$A \in \mathbf{A}$ & $A_1$ & $A_2$ & $A_3$ & $M_2$ \\ \hline
				$\mathbf{Z}_A$ & $C_2$ & $C_2, C_3, A_3$ & $A_2, C_3$ & $A_2, C_2, M_3$ \\
				\hline
			\end{tabular}
		\end{center}
		\caption{Intervention variables $A \in \mathbf{A}$ and induced dependences $\mathbf{Z}_A$ for the intervention in Fig. \ref{fig:gross-sg}}
		\label{tab:intervention}
	\end{table}
	
	\subsection{Identification Results}
	First, we show that the post-intervention $\G_{\mathbf{f_A}}$ is an SG.
	\begin{lem}
	\label{lem:g-fa}
		Given an SG $\mathcal{G}(\mathbf{V})$ and a segregation-preserving intervention $\mathbf{f_A}(\mathbf{Z_A})$, the post-intervention graph $\G_{\mathbf{f_A}}$ obtained via Procedure \ref{proc:gfa} is an SG.
	\end{lem}

	We now present the main result of this paper. This theorem provides sufficient conditions for the identification of the effects of policy interventions in SGs.
	\begin{thm}
		\label{thm:sg-policy-id}
		Let $\G(\mathbf{V} \cup \mathbf{H})$ be a causal LV-CG with $\mathbf{H}$ block-safe, and a topological order $\prec$. Fix disjoint $\mathbf{Y}, \mathbf{A} \subseteq \mathbf{V}$. Let $\mathbf{f_A}(\mathbf{Z_A})$ be a segregation preserving policy set. Let $\mathbf{Y}^{\star} \equiv \ant_{\G_{\mathbf{f_A}}}(\mathbf{Y}) \setminus \mathbf{A}$. Let $\G^d, \tilde{G}^d$ be the induced CADMGs on $\G_{\mathbf{f_A}}$ and $\G_{\mathbf{Y}^{\star}}$, and $\tilde{G}^b$ the induced CCG on $\G_{\mathbf{Y}^{\star}}$. Let $q(\mathbf{D}^{\star} | \pa^s_{\G_{\mathbf{f_A}}}(\mathbf{D}^{\star})) = \prod_{\mathbf{D} \in \G_{\mathbf{f_A}}} q(\mathbf{D} | \pa^s_{\G_{\mathbf{f_A}}}(\mathbf{D}))$, where $q(\mathbf{D} | \pa^s_{\G_{\mathbf{f_A}}}(\mathbf{D})) = \prod_{D \in \mathbf{D}} p(D | \mathbf{V}_{\prec D})$ if $\mathbf{D} \cap \mathbf{A} = \emptyset$ and $q = f_A(\mathbf{Z}_A)$ if $\mathbf{D} \cap \mathbf{A} \neq \emptyset$. $p(\mathbf{Y}(\mathbf{f_A}(\mathbf{Z_A})))$ is identified in $\G$ if and only if $p(\mathbf{Y}^{\star}(\mathbf{a}))$ is identified in $\G$ for the unrestricted class of policies. If identified,  $p(\mathbf{Y}(\mathbf{f_A}(\mathbf{Z_A}))) = $
		\begin{equation}
		\label{eq:sg-policy-id}
			\begin{aligned}
			\sum_{\{\mathbf{Y}^{\star} \cup \mathbf{A}\} \setminus \mathbf{Y}} &\bigg[ \prod_{\mathbf{B} \in \mathcal{B}(\tilde{\G}^b)} p^{\star}(\mathbf{B} | \pa_{\G_{\mathbf{f_A}}}(\mathbf{B})) \bigg]\\
			\times \bigg[\prod_{\mathbf{D} \in \mathcal{D}(\tilde{\G}^d)} &\phi_{\mathbf{D}^{\star} \setminus \mathbf{D}} (q(\mathbf{D}^{\star} | \pa^s_{\G_{\mathbf{f_A}}} (\mathbf{D}^{\star})); \G^d) \bigg] \bigg|_{\mathbf{A} = \tilde{\mathbf{a}}}
			\end{aligned}
		\end{equation}
		where (a) $\tilde{\mathbf{a}} = \{A = f_A(\mathbf{Z}_A) : A \in \pa_{\G_{\mathbf{f_A}}}(\mathbf{D}) \cap \mathbf{A} \}$ if $\pa_{\G_{\mathbf{f_A}}}(\mathbf{D}) \cap \mathbf{A} \neq \emptyset$ and $\tilde{\mathbf{a}}_{\mathbf{D}} = \emptyset$ otherwise, and (b) $p^{\star}$ is obtained by running Procedure \ref{proc:gibbs} over functions $g_{B_i}(B_{-i}, \pa_{\G_{\mathbf{f_A}}}(B_i), \epsilon_{B_i})$ where $g_{B_i} \in \mathbf{f_A}$ if $B_i \in \mathbf{A}$ and $g_{B_i}$ is given by the observed distribution if $B_i \not \in \mathbf{A}$\footnote{This distribution is identified from univariate terms but it cannot be obtained in closed-form.}.
	\end{thm}

	The outer sum over $\mathbf{A}$ is extraneous if $\mathbf{f_A}$ corresponds to a set of deterministic policies.

	\subsection{Estimands and Optimal Policy Selection}
	\label{sec:estimands}
	We now demonstrate how to obtain identified functionals via Eq. \ref{eq:sg-policy-id}. We describe identification of the effect on $\{Y_2, Y_3\}$ in Fig. \ref{fig:gross-sg}(a) of the intervention in Table \ref{tab:intervention}, and then give the functional for our elections example, Fig. \ref{fig:single-shot}(b), which we estimate in the next section.
	
	From Fig. \ref{fig:gross-sg}(a), we obtain $\G_{\mathbf{f_A}}$ in Fig. \ref{fig:gross-sg}(d) by applying the intervention detailed in Table \ref{tab:intervention}. In turn, from this post-intervention graph we observe that $\mathbf{Y}^{\star} = \ant_{\G_{\mathbf{f_A}}}(\mathbf{Y}) \setminus \mathbf{A} = \{C_2, C_3, M_3, Y_2, Y_3\}$ and obtain the induced subgraph $\G_{\mathbf{Y}^{\star}}$ in Fig. \ref{fig:gross-sg}(e).
	
	$\G_{\mathbf{Y}^{\star}}$ factorizes into kernels relating to district nodes and block nodes: $q_{\mathcal{D}}(C_1, A_1, M_1, Y_1, Y_2, Y_3|C_2, M_2, M_3)$ and $q_{\mathcal{B}}(M_2, M_3, A_2, A_3, C_2, C_3 | \emptyset)$. The block nodes factorize as a product of blocks, as in the first term of Eq. \ref{eq:sg-policy-id}. Separately, we must fix sets for each $\G_{\mathbf{Y}^{\star}}$ district $\{\{M_3\}, \{Y_2, Y_3\}\}$ in $q_{\mathcal{D}}$. This yields the functional (full derivation in the supplement) for $p(\{Y_2, Y_3\}(\mathbf{f_A}))$:
	\begin{align*}
		\sum_{\{A_1, A_2, A_3, M_2, M_3, C_2, C_3\}} \!\!\!\!\!\!\!\!\!\!\!\!\!\!\!\!\!\!\!\!\!\!p^{\star}(A_2, A_3 | C_2, C_3) &p^{\star}(M_2, M_3 | A_2, A_3, C_2)\\
		\times  p(Y_2, Y_3 | Y_1, A_1, M_1, &M_3, C_1, C_2)p^{\star}(C_2, C_3)
	\end{align*}
	
	Similarly, we consider the effect on $Y_l$ of intervening with a policy $f_{A_l}(C_l)$ in our electoral example, Fig. \ref{fig:single-shot} (b). $f_{A_l}(C_l)$ corresponds to a myopic strategy in which the candidate makes decisions based only on their own covariates. Applying Eq. \ref{eq:sg-policy-id}, $p(Y_l(f_{A_l}(C_l, C_r))) =$
	\begin{equation}
	\label{eq:estim-functional}
		\begin{aligned}
		\sum_{C_l, C_r, A_r, Y_r} &p(A_r | C_l, C_r) p(C_l) p(C_r)\\
		&\times p(Y_l, Y_r | C_l, C_r, A_r, f_{A_l}(C_l, C_r))
		\end{aligned}
	\end{equation}
	To choose an optimal action for the left candidate, we select $f_{A_l}(C_l)$ from a set of candidate policies $\mathcal{F}_{A_l}(C_l)$:
	\begin{align*}
	f_{A_l}(C_l, C_r) \,\,\, =\!\!\!\!\!\!\!\!\!\!\!\! \argmax_{\tilde{f}_{A_l}(C_l, C_r) \in \mathcal{F}_{A_l}(C_l, C_r)}\!\!\!\!\!\!\!\!\!\!\!\! p(Y_l(\tilde{f}_{A_l}(C_l, C_r)))
	\end{align*}

	\section{Estimation}
	\label{sec:experiments}
	\begin{figure*}[t]
		\begin{center}
			\subfloat[]{
				\includegraphics[width=1.95in]{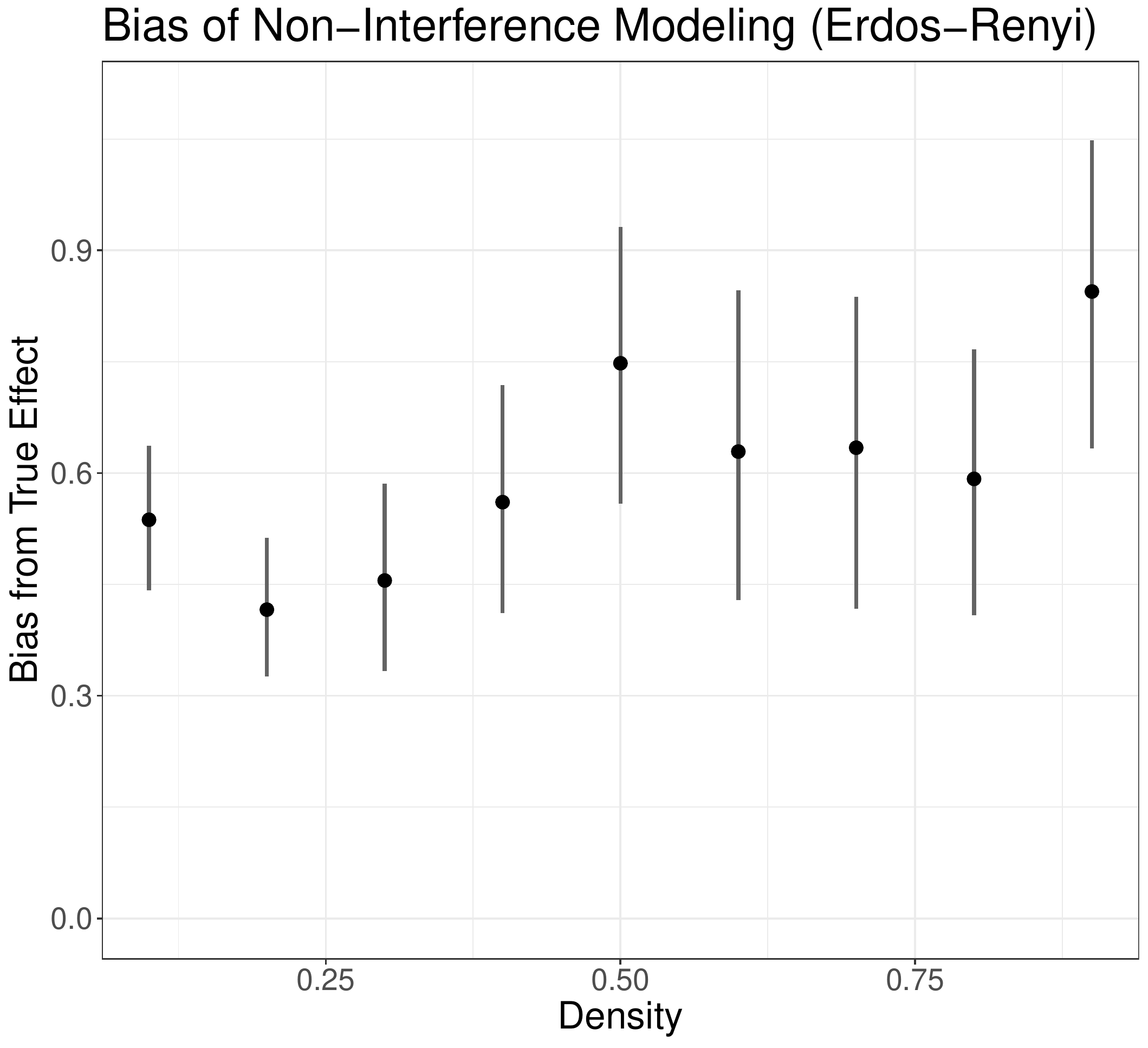}
				\label{fig:bias}
			}
			\hspace{2.5cm}
			\subfloat[]{\includegraphics[width=1.95in]{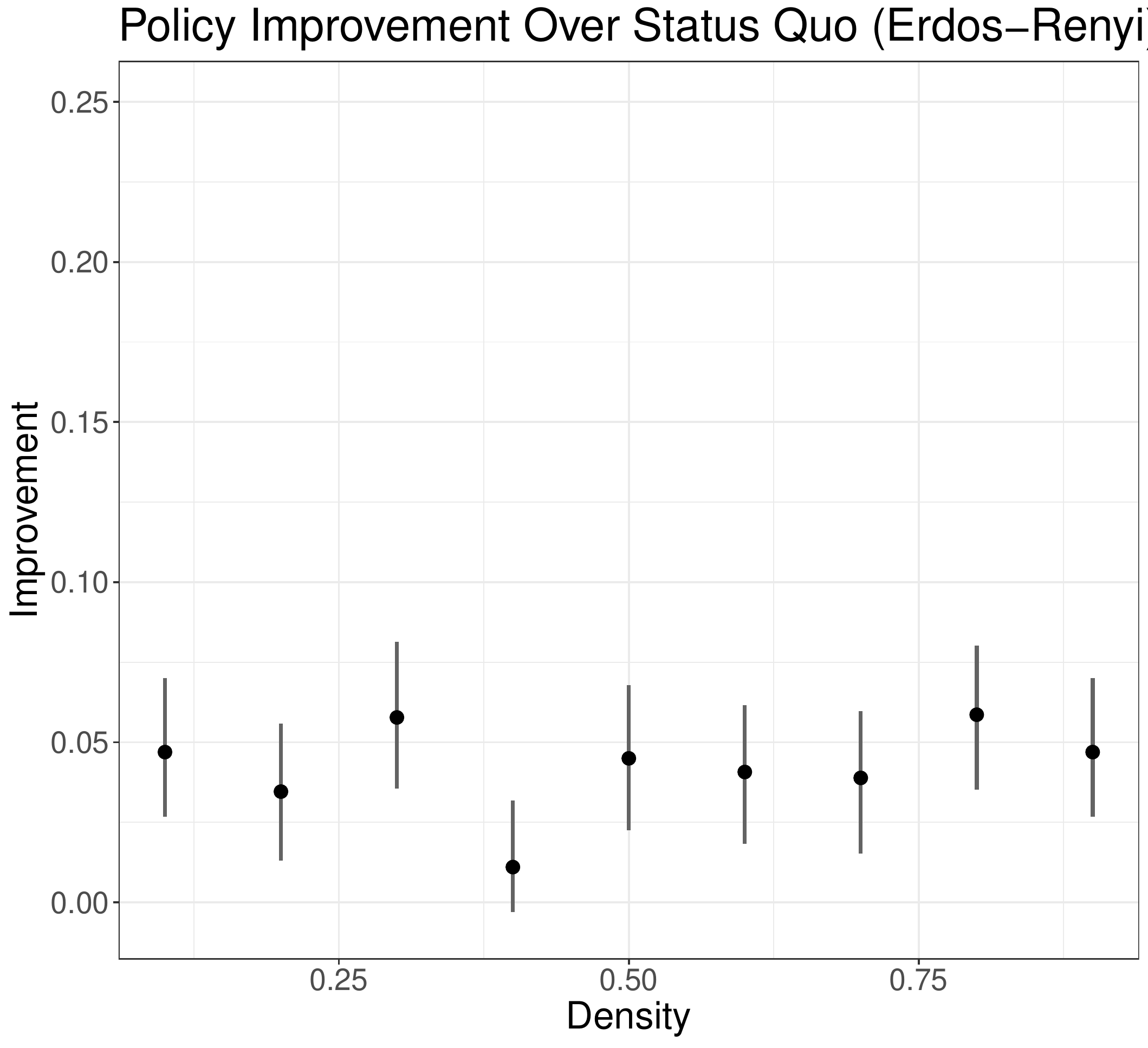} \label{fig:results}}
			\caption{(\ref{fig:bias}) Bias of estimates obtained using single-unit modeling, ignoring interference. The presence of bias suggests ignoring interference is highly problematic. (\ref{fig:results}) Difference in expected outcomes between an optimized strategy and the status quo. We analyze several network densities to demonstrate the generality of this approach.}
		\end{center}
	\end{figure*}
	We now demonstrate how functionals identified by Eq. \ref{eq:sg-policy-id} can be estimated from observed data. Specifically, we seek optimal $f_A(\mathbf{C})$'s for versions of the functional in Eq. \ref{eq:estim-functional}. To do so, we fit nuisance models and utilize the plug-in principle to perform indirect Q-learning for policy optimization. This approach yields consistent estimates of the optimized outcome under regularity conditions, assuming correctly specified nuisance models \citep{chakraborty2013statistical}.

	For our experiments we first generate 10-node network graphs according to one of three widely-used network generators: \citet{erdHos1960evolution}, \citet{watts1998collective}, and \citet{albert2002statistical}. In-unit and cross-unit structures are identical to the 2-node graph in Fig. \ref{fig:single-shot}(b). We then generate data for each $C$, $A$, and $Y$ using log-linear models, with $C \in [0,1]^3$ and $A, Y \in [0, 1]$.
	We use Gibbs sampling to approximate undirected edges between $Y$'s \citep{tchetgen2017auto}. We defer parametric specifications of our data generating process to the supplement. We assume \emph{partial} interference: we generate $1000$ samples of each network topology and use these to fit nuisance models. We run the following experiments by obtaining $1,000$ bootstrap replications of the generated data and calculating a $95\%$ confidence interval of the relevant effect:
	
	1. \textit{Bias from incorrectly assuming iid.}
	As a demonstration of the importance of using interference-aware modeling, we consider performing \emph{node} interventions on each $A_i$ obtained from our Erd\H{o}s-R\'enyi samples, setting $A_i$ to $1$ and $0$. We estimate the average causal effect (ACE) of these node interventions ($E[Y_i(1) - Y_i(0)]$) using models implied by ID (Table \ref{tab:id-simplification}, row three), which provides sound functionals when data \emph{are} iid, as well as models implied by the SG ID algorithm (Table \ref{tab:id-simplification}, row four) which respect the dependent nature of the data. We treat the latter models as `ground truth' and calculate the bias of the ACE induced by inappropriately assuming data are iid. These results are given in Fig. \ref{fig:bias}. Observing that bias is universally bounded away from $0$ in these results, it's clear that it's imperative to respect network dependence in causal modeling.

	2. \textit{Benefit of optimizing interventions.} Here we demonstrate the efficacy of policy interventions for picking tailored interventions that optimize a subject's outcome, by estimating the 10-unit version of the identified functional in Eq. \ref{eq:estim-functional}. From our generated samples, we fit logistic regression models for $E[Y_{-i} | \mathbf{A}, \mathbf{C}]$ and $E[Y_i | \mathbf{A}, \mathbf{C}, Y_{-i}]$, where $i$ denotes the unit we wish to optimize for. This ensures the necessary consistency properties for indirect Q-learning. Models for $p(\mathbf{A}, \mathbf{C})$ are estimated using the empirical distribution.
	
	For each sample we estimate the effect of intervening with a policy $f_{A_i}(C_i) \in \mathcal{F}_{A_i}(C_i) = \{|C_i|^{-1} \sum_{j \in [|C_i|]} k_j C_{ij} : k_j \in \mathbb{R}\}$ (i.e. $\mathcal{F}_{A_i}$ is the set of means of linear combinations of $C_i$'s components). We choose $\mathbf{k}$ to maximize $Y_i$ subject to the constraint that values of $A_i$ and $Y_i$ must remain in $[0,1]$. We report the \emph{difference} between the optimized and observed (`status quo') $Y_i$'s. The results for the  Erd\H{o}s-R\'enyi generator can be found in Fig. \ref{fig:results}. Results for the other generators can be found in the supplementary material.
	Since $Y$ is binary, an expected difference of $.05$ corresponds to a 5.0\% increase in $Y$ over the status quo. Fig. \ref{fig:results} demonstrates that the proposed approach virtually guarantees an improved outcome over the status quo.

	\section{Conclusion}
	In this paper we discussed identification of policy intervention effects in the interference setting. We characterized interpretations of possible interventions and gave criteria for identifying their effects in latent-variable causal chain graph models. Further, we demonstrated estimation via a simulation study. Future directions include exploring the intersection of policies and interference, and game theory, and developing robust estimation strategies for this setting.

	\clearpage
	
	\subsection*{Acknowledgements}
	The first author would like to thank the Adobe Research Internship program, and Sridhar Mahadevan, for supporting this work.
	The third author would like to thank the following organizations for supporting this work: the American Institute of Mathematics SQuaRE program, National Institutes of Health grant R01 AI127271-01A1, Office of Naval Research grant N00014-18-1-2760, and Defense Advanced Research Progress Administration grant under contract HR0011-18-C-0049. The
	content of the information in this paper does not necessarily reflect the position or the policy of the Government, and no official endorsement should be inferred.
	\bibliographystyle{abbrvnat}
	\bibliography{references}
	
	\onecolumn
	\appendix


\section{Graphical Models Background}
\subsection{Statistical Graphical Models}
Chain graphs and their submodels were originally conceived as \emph{statistical} models over random variables, encoding conditional independence constraints in their factorization. For instance, a DAG $\mathcal{G}(\mathbf{V})$ represents the set of joint distributions over $\mathbf{V}$ which factorize according to:
\begin{align}
\label{eq:dag-fact}
p(\mathbf{V}) = \prod_{V \in \mathbf{V}} p(V | \pa_{\mathcal{G}}(V))
\end{align}
Similarly, an MRF $\mathcal{G}(\mathbf{V})$ represents the set of distributions over $\mathbf{V}$ which factorize according to the factorization:
\begin{align*}
p(\mathbf{V}) = Z^{-1} \prod_{\mathbf{C} \in \mathcal{C}(\mathcal{G}_{U})} \phi_{\mathbf{C}}(\mathbf{C}),
\end{align*}
where $Z$ is a normalizing constant, $\mathcal{C}(\mathcal{G})$ denotes the set of cliques in $\mathcal{G}$, and $\phi_{\mathbf{C}}$ is an arbitrary function over $\mathbf{C}$ known as a clique potential \cite{lauritzen1996graphical}.

CGs merge these notions by allowing for both directed and undirected edges. Like DAGs and MRFs, a CG $\mathcal{G}(\mathbf{V})$ represents the set of distributions over $\mathbf{V}$ that factorize according to the two-level factorization:
\begin{align}
p(\mathbf{V}) &= \prod_{\mathbf{B} \in \mathcal{B}(\mathcal{G})} p(\mathbf{B} | pa_{\mathcal{G}}(\mathbf{B}))\\
p(\mathbf{B} | pa_{\mathcal{G}}(\mathbf{B})) &= Z(pa_{\mathcal{G}}(\mathbf{B}))^{-1} \prod_{\mathbf{C} \in \mathcal{C}^{\star}} \phi_{\mathbf{C}}(\mathbf{C}),
\end{align}
where $\mathcal{C}^{\star} = \{\mathbf{C} \in \mathcal{C}((\mathcal{G}_{\mathbf{B} \cup pa_{\mathcal{G}} (\mathbf{B}) })^a) : \mathbf{C} \not \subseteq pa_{\mathcal{G}}(\mathbf{B})\}$, the set of cliques that intersect $\mathbf{B}$ in the augmented graph \cite{lauritzen1996graphical} on $\mathbf{B}$ and $\pa_{\G}(\mathbf{B})$. An augmented graph $\G^a$ is obtained from $\G$ by making any edges in $\G$ undirected and adding undirected edges between each $V \in \pa_{\G}(\mathbf{B})$ for all $\mathbf{B} \in \mathcal{B}(\G)$.

Throughout the paper, we will assume that all probability distributions have full support.

\subsection{Causal Graphical Models}
In contrast to their statistical analogues, causal DAGs \cite{pearl2000causality} and causal CGs \cite{lauritzen2002chain} represent distributions over \emph{counterfactual} variables. For $Y \in \mathbf{V}$ and $\mathbf{A} \subseteq \mathbf{V} \setminus V$, the counterfactual $Y(\mathbf{a})$ denotes the value of $Y$ under the hypothetical scenario in which $\mathbf{A}$ is set to $\mathbf{a}$ via a \emph{node} (or \emph{atomic}) intervention \cite{pearl2000causality}. 

In this paper, we will assume Pearl's functional model. For a DAG $\G(\mathbf{V})$, counterfactuals $V(\mathbf{a})$ are determined by structural equations $f_{V}(\mathbf{a}, \epsilon_V)$, which remain invariant under an intervention $\mathbf{a}$ with $\epsilon_V$ denoting randomness in the causal process. These \emph{one step ahead} counterfactuals can be used to define all variables in the model via \emph{recursive substitution}. For $\mathbf{A} \subseteq \mathbf{V} \setminus \{V\}$:
\begin{align*}
V(\mathbf{a}) \equiv V(\mathbf{a}_{\pa_{\G}(V)}, \{W(\mathbf{a}) : W \in \pa_{\G}(V) \setminus \mathbf{A}\}),
\end{align*}
where $\mathbf{a}$ lies in the state space of $\mathbf{A}$.

A parameter is said to be \emph{identifiable} in a model when it can be expressed as a function of observed data. In a DAG or CG $\G$ with all variables $\mathbf{V}_D$ or $\mathbf{V}_C$, all counterfactuals arising from node interventions are identified by the \emph{g-formula} \cite{robins1986new} and \emph{chain graph g-formula} \cite{lauritzen2002chain} respectively:
\begin{align}
\label{eq:g-formula}
p(\mathbf{V}_D(\mathbf{a})) &= \prod_{V \in \mathbf{V}_D \setminus \mathbf{A}} p(V | \pa_{\G}(V))|_{\mathbf{A} = \mathbf{a}}\\
\label{eq:cg-g-formula}
p(\mathbf{V}_C(\mathbf{a})) &= \prod_{\mathbf{B} \in \mathcal{B}(\G)} p(\mathbf{B} \setminus \mathbf{A} | \pa_{\G}(\mathbf{B}), \mathbf{B} \cap \mathbf{A})|_{\mathbf{A} = \mathbf{a}}
\end{align}

\section{Background: Identification in Causal Graphical Models}
In this section, we discuss the state of latent-variable identification theory in causal graphical models. These advancements culminate with sound and complete algorithms for identification in the presence of latent variables in segregated graphs and identification of policy intervention effects in latent variable DAGs. The current work bridges these literatures.

\subsection{The Nested Markov Factorization: Re-expressing the ID Algorithm}
\cite{tian2002general} gave a general condition for determining identifiability of node interventions in latent-variable DAGs. \cite{shpitser2006identification} re-expressed Tian's condition as a concise algorithm, known as `ID', and proved that it is complete. Recently, \cite{richardson2017nested} re-expressed the algorithm in \cite{shpitser2006identification} in terms of a modified nested factorization, similar to the g-formula in a type of mixed graph.

While the substance of the identification approach in \cite{richardson2017nested} is identical to that in \cite{shpitser2006identification} and \cite{tian2002general}, the fixing operator they present enables a compact representation of existing identification theory and makes clear the connection between the ID algorithm and Robins' g-formula (Eq. \ref{eq:g-formula}), which is itself a modified factorization on DAGs. For these reasons, we make use of this framework for the SG policy identification extension that we present in this work. This re-formulation relies on several important concepts which we describe below.

\subsubsection{Latent Projections}
Rather than considering latent-variable DAGs explicitly, \cite{richardson2017nested} considers a class of models known as acyclic directed mixed graphs (ADMGs). An ADMG contains directed and bi-directed edges and represents and equivalence class of latent-variable DAGs. Given a latent-variable DAG $\G(\mathbf{V} \cup \mathbf{H})$, where $\mathbf{V}$ is observed and $\mathbf{H}$ is latent, we can obtain the corresponding ADMG $\G'(\mathbf{V})$ via a latent projection operation: edges $A \rightarrow B$ in $\G$ are maintained in $\G'$; additionally, $\G'$ has an edge $A \rightarrow B$ for any directed path $A \rightarrow \dots \rightarrow B$ where the intermediate nodes are all in $\mathbf{H}$, and $\G'$ has an edge $A \leftrightarrow B$ if there exists a path $A \leftarrow \dots \rightarrow B$ in $\G$ with all intermediate nodes in $\mathbf{H}$ and no consecutive edges $\rightarrow H \leftarrow$ for $H \in \mathbf{H}$. We can also define conditional ADMGs (CADMGs) which partition nodes into sets of random variables $\mathbf{V}$ and fixed variables $\mathbf{W}$. In a CADMG $\G(\mathbf{V}, \mathbf{W})$, the variables $\mathbf{W}$ have no incoming edges. An ADMG $\G(\mathbf{V})$ is also trivially a CADMG with $\mathbf{W} = \emptyset$.

As described above, we use segregated graphs as the chain graph analogue of ADMGs. In the formulation we use, SGs represent an equivalence class of latent variable chain graphs, defined in a way that maintains their causal interpretation. A latent variable chain graph $\G(\mathbf{V} \cup \mathbf{H})$ is \emph{block-safe} \cite{sherman2018identification} if no $V \in \mathbf{V}$ has an incident edge from a latent variable (i.e., $H \rightarrow V$ for $H \in \mathbf{H}$ is forbidden) and no latent variable $H \in \mathbf{H}$ has an incident undirected edge. A block-safe latent variable chain graph can be represented with a segregated graph via the same latent projection operation described above.

\subsubsection{Kernels and Fixing}
Whereas DAGs and CGs factorize as a product of conditional probability distributions, ADMGs and SGs factorize as a product of \emph{kernels} \cite{lauritzen1996graphical}. Again, following the notation in \cite{sherman2018identification}, a kernel $q_{\mathbf{V}}(\mathbf{V} | \mathbf{W})$ is a function that maps values of $\mathbf{W}$ to densities on $\mathbf{V}$ with $\sum_{\mathbf{v} \in \mathbf{V}} q_{\mathbf{V}}(\mathbf{v} | \mathbf{w}) = 1$ for each possible realization $\mathbf{w}$. As with probability distributions, for some $\mathbf{A} \subseteq \mathbf{V}$, conditioning and marginalization in kernels are defined as follows:
\begin{align*}
q(\mathbf{A} | \mathbf{W}) &\equiv \sum_{\mathbf{V} \setminus \mathbf{A}} q(\mathbf{V} | \mathbf{W})\\
q(\mathbf{V} \setminus \mathbf{A} | \mathbf{A}, \mathbf{W}) &\equiv \frac{q(\mathbf{V} | \mathbf{W})}{q(\mathbf{A} | \mathbf{W})}.
\end{align*}

The notion of fixing variables is closely tied to kernels. In a CADMG $\G(\mathbf{V})$, a variable $V$ is fixable if there does not exist both a bi-directed path and a directed path to some $V' \in \mathbf{V}$, or concisely, $\de_{\G}(V) \cap \dis_{\G}(V) = \emptyset$. In a DAG $\G$ with a corresponding probability distribution $p(\mathbf{V})$, fixing $V$ corresponds to applying the g-formula to obtain a new distribution $p(\mathbf{V} \setminus V)$ and a new graph $\G'$. For a CADMG $\G(\mathbf{V}, \mathbf{W})$ with a corresponding kernel $q(\mathbf{V} | \mathbf{W})$, \cite{richardson2017nested} defines a similar operation for fixing $V$ in $\G$, denoted $\phi_V(\G)$. This operator yields a new kernel and a new CADMG $\G'(\mathbf{V} \setminus \{V\}, \mathbf{W} \cup \{V\})$. In this graph, all bi-directed and directed edges \emph{into} $V$ are removed. The operator also yields a new kernel:
\begin{align*}
q'(\mathbf{V} \setminus \{V\} | \mathbf{W} \cup \{V\}) \equiv \frac{q(\mathbf{V} | \mathbf{W})}{q(\mathbf{V} | \nd_{\G}(V), \mathbf{W})}.
\end{align*}
Since the fixing operation generalizes the g-formula, it's probabilistic interpretation varies -- acting as marginalization, conditioning, and sometimes neither -- depending on the characteristics of the variable being fixed relative to the kernel it is being fixed in.

\subsubsection{Reachability, the Nested Factorization, and ID}
We can extend the notion of fixability to sets of variables $\mathbf{S} \subseteq \mathbf{V}$ in a CADMG $\G$. If it is possible to find a sequence $S_1, S_2, \dots$ of the variables in $\mathbf{S}$ such that $S_1$ is fixable in $\G$, $S_2$ is fixable in $\phi_S(\G)$ and so on, then $\mathbf{S}$ is fixable and $\mathbf{V} \setminus \mathbf{S}$ is said to be \emph{reachable} in $\G$.

It was shown in \cite{richardson2017nested} and \cite{sherman2018identification} that all valid fixing sequences for $\mathbf{S}$ in a CADMG $\G(\mathbf{V}, \mathbf{W})$ yield the same resulting CADMG $\G(\mathbf{V} \setminus \mathbf{S}, \mathbf{W} \cup \mathbf{S})$ and analogously for the kernel obtained by fixing $\mathbf{S}$ in $q(\mathbf{V} | \mathbf{W})$. The fixing operator can therefore be defined for sets as it was for singleton variables: $\phi_{\mathbf{S}}$. A CADMG $\G(\mathbf{V}, \mathbf{W})$ is said to satisfy the nested Markov factorization if for every fixable $\mathbf{S}$
\begin{align*}
\phi_{\mathbf{S}}(q(\mathbf{V} | \mathbf{W}); \G) = \prod_{\mathbf{D} \in \mathcal{D}(\phi_{\mathbf{S}}(\G))} \phi_{\mathbf{V} \setminus \mathbf{D}}(q(\mathbf{V} | \mathbf{W}); \G)
\end{align*}
\cite{richardson2017nested} showed that $p(\mathbf{V} \cup \mathbf{H})$ satisfies the above factorization for a DAG $\G(\mathbf{V} \cup \mathbf{H})$ then $p(\mathbf{V})$ satisfies the factorization for the corresponding ADMG $\G(\mathbf{V})$. An analogous result for SGs was shown in \cite{sherman2018identification}, which we will discuss below.

This notation permits a reformulation of the ID algorithm as a one line formula, proven in \cite{richardson2017nested} to be identical to the algorithm in \cite{shpitser2006identification}: Let $\mathbf{Y}, \mathbf{A}$ be disjoint subsets of $\mathbf{V}$ in an ADMG $\G(\mathbf{V})$. Let $\mathbf{Y}^{\star} = \an_{\G_{\mathbf{V} \setminus \mathbf{A}}}(\mathbf{Y})$. The intervention $p(\mathbf{Y} | \text{do}(\mathbf{a}))$ is identified in $\G$ if and only if every set (district) $\mathbf{D} \in \mathcal{D}(\G_{\mathbf{Y}^{\star}})$ is reachable and, if identification holds, then
\begin{align}
\label{eq:vanilla-id}
p(\mathbf{Y} | \text{do}(\mathbf{a})) = \sum_{\mathbf{Y}^{\star} \setminus \mathbf{Y}} \prod_{\mathbf{D} \in \mathcal{D}(\G_{\mathbf{Y}^{\star}})} \phi_{\mathbf{V} \setminus \mathbf{D}}(p(\mathbf{V}); \G) |_{\mathbf{A} = \mathbf{a}}.
\end{align}

\subsection{Identification in Segregated Graphs}
\subsubsection{The Segregated Factorization}
Building off the nested factorization for ADMGs and the chain graph factorization, we can define the segregated factorization of an SG \cite{sherman2018identification}. Recall that the variables in an SG $\G$ can be grouped into those that lie in a non-trivial block which we denote $\mathbf{B}^{\star} = \cup_{\mathbf{B} \in \mathcal{B}^{nt}(\G)} \mathbf{B}$, and those that don't, which we denote $\mathbf{D}^{\star} = \cup_{\mathbf{D} \in \mathcal{D}(\G)} \mathbf{D}$. 

We can factorize an SG as the product of two kernels. The first kernel corresponds to a conditional chain graph (CCG) $\G(\mathbf{V}, \mathbf{W})$ where, as in CADMGs, $\mathbf{V}$ are random nodes and $\mathbf{W}$ are fixed. A kernel $q(\mathbf{V} | \mathbf{W})$ is said to be Markov relative to a CCG $\G$ if it satisfies Eq. \ref{eq:cg-fact} with the following modification to the outer factorization
\begin{align*}
q(\mathbf{V} | \mathbf{W}) = Z(\mathbf{W})^{-1} \prod_{\mathbf{B} \in \mathcal{B}(\G)} q(\mathbf{B} | \pa_{\G}(\mathbf{B})),
\end{align*}
and a similar replacement of $p(\mathbf{B} | \pa_{\G}(\mathbf{B}))$ with $q$ in the inner factorization.	We will denote the CCG obtained from a SG by $\G^b$ with $\mathbf{V}$ corresponding to $\mathbf{B}^{\star}$ and $\mathbf{W}$ to $\pa_{\G}^s(\mathbf{B}^{\star})$. $\G^b$ contains edges between each node in $\mathbf{B}^{\star}$ that exists in $\G$ as well as those between $\pa_{\G}^s(\mathbf{B}^{\star})$ in $\G$. 

The second kernel corresponds to a CADMG which we will denote $\G^d$ with random nodes $\mathbf{D}^{\star}$ and fixed nodes $\pa_{\G}^s(\mathbf{D}^{\star})$. $\G^d$ contains all edges between $\mathbf{D}^{\star}$ that are present in $\G$ as well as the edges between $\pa_{\G}^s(\mathbf{D}^{\star})$ and $\mathbf{D}^{\star}$ in $\G$. 

If each of these kernels adheres to the factorization of the respective conditional graph, then $p(\mathbf{V})$ is obeys the segregated factorization. Specifically, $p(V)$ satisfies the segregated factorization if $q(\mathbf{D}^{\star} | \pa_{\G}^s(\mathbf{D}^{\star})$ satisfies the nested factorization and $q(\mathbf{B}^{\star} | \pa_{\G}^s(\mathbf{B}^{\star}))$ satisfies the CCG factorization.

\subsubsection{The Segregated Graph ID Algorithm}
Using the above extension of the nested factorization, we can now describe the extension to the ID algorithm, expressed using the fixing operator $\phi$ \cite{sherman2018identification}: for a block safe segregated graph $\G(\mathbf{V})$, fix disjoint $\mathbf{Y}, \mathbf{A} \subseteq \mathbf{V}$. Similar to above, let $\mathbf{Y}^{\star} = \ant_{\G_{\mathbf{V} \setminus \mathbf{A}}} (\mathbf{Y})$. Define $\tilde{\G}^d$ and $\tilde{\G}^b$ to be the CADMG and CCG respectively obtained from $\G_{\mathbf{Y}^{\star}}$ and 
\begin{align*}
q(\mathbf{D}^{\star}) = \frac{p(\mathbf{V})}{\prod_{\mathbf{B} \in \mathcal{B}^{nt}(\G)} p(\mathbf{B} | \pa_{\G}(\mathbf{B}))}.
\end{align*}
We then have $p(\mathbf{Y} | \text{do}(\mathbf{a}))$ is identified in $\G$ if and only if $\mathcal{D}(\tilde{\G}^d)$ is reachable in $\G^d$ and, if it is identified, then it is equal to
\begin{equation}
\label{eq:sg-id}
\begin{aligned}
\sum_{\mathbf{Y}^{\star} \setminus \mathbf{Y}} &\bigg[ \prod_{\mathbf{D} \in \mathcal{D}(\tilde{\G}^d)} \phi_{\mathbf{D}^{\star} \setminus \mathbf{D}} (q(\mathbf{D}^{\star} | \pa_{\G}(\mathbf{D}^{\star})); \G^d) \bigg]\\
\times &\bigg[ \prod_{\mathbf{B} \in \mathcal{B}(\tilde{\G}^b)} p(\mathbf{B} \setminus \mathbf{A} | \pa_{\G_{\mathbf{Y}^{\star}}}(\mathbf{B}), \mathbf{B} \cap \mathbf{A}) \bigg] \bigg|_{\mathbf{A} = \mathbf{a}}.
\end{aligned}
\end{equation}

\subsection{Policy Interventions in ADMGs}
Extending node interventions, \cite{tian2008identifying} proposed a framework for setting an intervention node in a DAG to a policy, a function of variables preceding it in the graph. Formally, for a DAG $\mathcal{G}(\mathbf{V})$ with a topological ordering $\prec$ on $\mathbf{V}$ and an intervention set $\mathbf{A} \subseteq \mathbf{V}$, let $\mathbf{f}_{\mathbf{A}}$ be the set of policies $f_A$ corresponding to each node $A$ in $\mathbf{A}$. Each $f_A$ is a function of some set $\mathbf{W}_A \subseteq \mathbf{V}_{\prec A}$ such that it maps the state space of $\mathbf{W}_{A}$ to the state space of $A$. Graphically, intervening with $f_A$ corresponds to removing all edges \emph{into} $A$ in $\mathcal{G}$ and adding in edges from $\mathbf{W}_A$ to $A$, yielding a new graph $\mathcal{G}_{\mathbf{f_A}}$.

For an intervention of this type, we can define a counterfactual $Y(\mathbf{f}_{\mathbf{A}})$ for $Y \in \mathbf{V}$ analogously to node interventions via recursive substitution:
\begin{align*}
Y(\{f_A( \mathbf{W}_{\mathbf{A}}( \mathbf{f}_{\mathbf{A}} ) ) | A \in \pa_{\mathcal{G}}(Y) \cap \mathbf{A}\}, \{\pa_{\G}(Y) \setminus \mathbf{A}\} (\mathbf{f}_{\mathbf{A}}))
\end{align*}

This implies a policy-analogue \cite{tian2008identifying} to the g-formula for $p(\{\mathbf{V} \setminus \mathbf{A}\} (\mathbf{f}_{\mathbf{A}}))$:
\begin{align*}
\prod_{V \in \mathbf{V} \setminus \mathbf{A}} p(V | \{f_A(\mathbf{W}_A) : A \in \mathbf{A} \cap \pa_{\G}(V)\}, \pa_{\G}(V) \setminus \mathbf{A})
\end{align*}
An extension of these ideas to latent-variable DAGs was given in \cite{shpitser2018identification}. Following the Richardson re-expression of the ID algorithm, for an ADMG $\G$, the post-intervention graph $\G_{\mathbf{f_A}}$ is obtained in the same way as the fully observed case: by removing edges into $\mathbf{A}$ and adding edges from $\mathbf{W}_A$ to $\mathbf{A}$. Similarly, \cite{shpitser2018identification} defines $\mathbf{Y}^{\star} \equiv \an_{\G_{\mathbf{f_A}}}(\mathbf{Y}) \setminus \mathbf{A}$. This leads to a policy-analogue of Eq. \ref{eq:vanilla-id}: $p(\mathbf{Y}(\mathbf{f_A}))$ is identified in $\G$ if and only if $p(\mathbf{Y}^{\star}(\mathbf{a}))$ is identified in $\G$; if it is identified then
\begin{align*}
p(\mathbf{Y}(\mathbf{f_A})) = \sum_{(\mathbf{Y}^{\star} \cup \mathbf{A}) \setminus \mathbf{Y}} \prod_{\mathbf{D} \in \mathcal{D}(\G_{\mathbf{Y}^{\star}})} \phi_{\mathbf{V} \setminus \mathbf{D}}(p(\mathbf{V}); \G)|_{\tilde{\mathbf{a}}_{\pa_{\G}^s(\mathbf{D}) \cap \mathbf{A}}}
\end{align*}
where $\tilde{\mathbf{a}}_{\pa_{\G}^s(\mathbf{D}) \cap \mathbf{A}} = \{A = f_A(\mathbf{W}_A) | A \in \pa_{\G}(\mathbf{D}) \cap \mathbf{A}\}$ if $\pa_{\G}(\mathbf{D}) \cap \mathbf{A} \neq \emptyset$ and $\tilde{\mathbf{a}} = \emptyset$ otherwise.	
	
\section{Proofs}
\begin{lema}{\ref{lem:g-fa}}
	Given a segregated graph $\mathcal{G}(\mathbf{V})$ and a segregation-preserving policy intervention $\mathbf{f_A}(\mathbf{Z_A})$, the post-intervention graph $\G_{\mathbf{f_A}}$ obtained via Procedure \ref{proc:gfa} is a segregated graph.
\end{lema}

\begin{prf}
	In order for $\G_{\mathbf{f_A}}$ to be a segregated graph, it must not have a node with both an incident bi-directed and undirected edge (the `segregation' property) and it must not have any partially directed cycles (the `chain' property).
	
	We first show that $\G_{\mathbf{f_A}}$ satisfies the segregation property. First we consider edges that appear in both $\G$ and $\G_{\mathbf{f_A}}$ (potentially with a modified functional form). Since we do not add any $\leftrightarrow$ edges when constructing $\G_{\mathbf{f_A}}$, and since we assumed $\G$ is a segregated graph, these edges are all incident to nodes that do not also have incident directed edges. 
	
	We can therefore restrict attention to undirected edges that were newly created when constructing $\G_{\mathbf{f_A}}$. These edges correspond to connecting two previously unconnected nodes. This requires intervening on both end points, which entails removing all incident $\leftrightarrow$ edges, as described in Procedure \ref{proc:gfa}. This accounts for all possible undirected edges. In particular, we cannot convert a directed edge $X \rightarrow Y$ to an undirected edge $X - Y$: this would require intervening on $X$ with $f_X(\mathbf{Z}_X)$ where $Y \in \mathbf{Z}_X$ which violates our construction that $\mathbf{Z} \subseteq \mathbf{V} \setminus \overline{\ext}_{\G}(X)$.
	
	Since no undirected edge is incident to a node that also has an incident bi-directed edge, $\G_{\mathbf{f_A}}$ satisfies the segregation property.
	
	We now show that $\G_{\mathbf{f_A}}$ satisfies the chain property. We argue by contradiction: suppose $\G_{\mathbf{f_A}}$ \emph{does} have a newly induced (relative to $\G$) partially directed cycle. Then, without loss of generality, one of the following sub-structures appears in $\G_{\mathbf{f_A}}$ but not in $\G$: (1) $W \rightarrow X \rightarrow Y \rightarrow W$, (2) $W \rightarrow X - Y \rightarrow W$, or (3) $W \rightarrow X - Y - W$.
	
	Sub-structure (1) contradicts our assumption that $\mathbf{f_A}$ is segregation-preserving. Specifically, we have that $W \triangle X$ directly and $X \triangle W$ through $Y$, however $W \not \in \mathbf{Z}_Y$.
	
	In sub-structure (2), consider scenarios where two edges were present in $\G$ and we seek to add the third edge. When adding either the $W \rightarrow X$ or $Y \rightarrow W$ edge, we have that $W \triangle X$ and $X \triangle W$ (analogously for $W, Y$) but $X \not \in \mathbf{Z}_W$ ($W \not \in \mathbf{Z}_Y$) which is a contradiction. Meanwhile, adding the $X - Y$ edge requires that $Y \in \mathbf{Z}_X$, however $Y \in \overline{\ext}_{\G}(X)$ which yields a contradiction. A similar argument involving $\triangle$ applies when only one of the three edges was present in $\G$ and we seek to add the other two.
	
	In sub-structure (3) a similar argument applies. Suppose we seek to add the $W \rightarrow X$ edge with the two undirected edges present. $X \triangle W$ in the post-intervention graph but it is not the case that $W \triangle X$, yielding a contradiction. Adding the $Y - X$ edge yields a contradiction since $X \in \overline{\ext}_{\G}(Y)$. Similarly, adding the $Y - W$ edge yields a contradiction since $Y \in \overline{\ext}_{\G}(W)$. Again, we can make a similar argument for adding two of the three edges.
	
	The above argument generalizes trivially to larger sub-structures in the graph (e.g., $4$-cycles) and so $\G_{\mathbf{f_A}}$ will not have any partially directed cycles. Since $\G_{\mathbf{f_A}}$ satisfies both the chain property and the segregation property, it is a segregated graph.
\end{prf}

\begin{thma}{\ref{thm:sg-policy-id}}
	Let $\G(\mathbf{V} \cup \mathbf{H})$ be a causal LV-CG with $\mathbf{H}$ block-safe, and a topological order $\prec$. Fix disjoint $\mathbf{Y}, \mathbf{A} \subseteq \mathbf{V}$. Let $\mathbf{f_A}(\mathbf{Z_A})$ be a segregation preserving policy set. Let $\mathbf{Y}^{\star} \equiv \ant_{\G_{\mathbf{f_A}}}(\mathbf{Y}) \setminus \mathbf{A}$. Let $\G^d, \tilde{G}^d$ be the induced CADMGs on $\G_{\mathbf{f_A}}$ and $\G_{\mathbf{Y}^{\star}}$, and $\tilde{G}^b$ the induced CCG on $\G_{\mathbf{Y}^{\star}}$. Let $q(\mathbf{D}^{\star} | \pa^s_{\G_{\mathbf{f_A}}}(\mathbf{D}^{\star})) = \prod_{\mathbf{D} \in \G_{\mathbf{f_A}}} q(\mathbf{D} | \pa^s_{\G_{\mathbf{f_A}}}(\mathbf{D}))$, where $q(\mathbf{D} | \pa^s_{\G_{\mathbf{f_A}}}(\mathbf{D})) = \prod_{D \in \mathbf{D}} p(D | \mathbf{V}_{\prec D})$ if $\mathbf{D} \cap \mathbf{A} = \emptyset$ and $q = f_A(\mathbf{Z}_A)$ if $\mathbf{D} \cap \mathbf{A} \neq \emptyset$. $p(\mathbf{Y}(\mathbf{f_A}(\mathbf{Z_A})))$ is identified in $\G$ if and only if $p(\mathbf{Y}^{\star}(\mathbf{a}))$ is identified in $\G$ for the unrestricted class of policies. If identified,  $p(\mathbf{Y}(\mathbf{f_A}(\mathbf{Z_A}))) = $
	\begin{equation}
	\label{eq:sg-policy-id-apx}
	\begin{aligned}
	\sum_{\{\mathbf{Y}^{\star} \cup \mathbf{A}\} \setminus \mathbf{Y}} &\bigg[ \prod_{\mathbf{B} \in \mathcal{B}(\tilde{\G}^b)} p^{\star}(\mathbf{B} | \pa_{\G_{\mathbf{f_A}}}(\mathbf{B})) \bigg]\\
	\times \bigg[\prod_{\mathbf{D} \in \mathcal{D}(\tilde{\G}^d)} &\phi_{\mathbf{D}^{\star} \setminus \mathbf{D}} (q(\mathbf{D}^{\star} | \pa^s_{\G_{\mathbf{f_A}}} (\mathbf{D}^{\star})); \G^d) \bigg] \bigg|_{\mathbf{A} = \tilde{\mathbf{a}}}
	\end{aligned}
	\end{equation}
	where (a) $\tilde{\mathbf{a}} = \{A = f_A(\mathbf{Z}_A) : A \in \pa_{\G_{\mathbf{f_A}}}(\mathbf{D}) \cap \mathbf{A} \}$ if $\pa_{\G_{\mathbf{f_A}}}(\mathbf{D}) \cap \mathbf{A} \neq \emptyset$ and $\tilde{\mathbf{a}}_{\mathbf{D}} = \emptyset$ otherwise, and (b) $p^{\star}$ is obtained by running Procedure \ref{proc:gibbs} over functions $g_{B_i}(B_{-i}, \pa_{\G_{\mathbf{f_A}}}(B_i), \epsilon_{B_i})$ where $g_{B_i} \in \mathbf{f_A}$ if $B_i \in \mathbf{A}$ and $g_{B_i}$ is given by the observed distribution if $B_i \not \in \mathbf{A}$\footnote{This distribution is identified from univariate terms but it cannot be obtained in closed-form.}.
\end{thma}

\begin{prf}
	We prove two subclaims.
	
	\textbf{Claim 1: The segregated graph policy ID formula, equation \ref{eq:sg-policy-id-apx}, is sound}
	
	We first note that each variable in $\mathbf{V} \cup \mathbf{H}$ is defined by a structural equation model. Since $\mathbf{f_A}$ is assumed to be segregation preserving, lemma \ref{lem:g-fa} implies that all variables in $\mathbf{H}$ have an unchanged structural equation in $\mathbf{f_A}$. Among $\mathbf{V}$ there exist two types of variables: those that have a symmetric functional dependence with another variable (i.e., for $V_i, V_j \in \mathbf{V}$ the structural equations $f_{V_i}, f_{V_j}$ are functions of each other), and those without symmetric dependence.
	
	We impose an ordering on the variables in $\G_{\mathbf{f_A}}$ in order of their dependence on other variables in the graph: we first evaluate variables $\mathbf{V} \in (\mathbf{V} \cup \mathbf{H})$ with structural equations that don't depend on other variables ($V \sim f_V(\epsilon_V)$) and then variables that are functions of those variables and so on. Following \cite{lauritzen2002chain}, groups of variables that have symmetrically dependent structural equations are chain components corresponding to $\mathcal{B}^{nt}(\G_{\mathbf{f_A}})$. Variables that do not exhibit symmetric dependence are trivial chain components. Our ordering therefore implies a DAG on chain components (it is acyclic aside from in-component cycles by lemma \ref{lem:g-fa}). 
	
	It's clear that for trivial chain components the functions $f_V$ immediately reach an equilibrium. We can normalize these functions, and write the margin over their corresponding variables as:
	\begin{align*}
	\prod_{V \in \mathbf{D} : \mathbf{D} \in \mathcal{D}(\G_{\mathbf{f_A}}(\mathbf{V} \cup \mathbf{H}))} p(V | \pa_{\G_{\mathbf{f_A}}(\mathbf{V} \cup \mathbf{H})}(V))|_{\mathbf{A} = \mathbf{f_A}}
	\end{align*}
	
	Now, for each non-trivial chain component $\mathbf{B}$, the structural equations for each constituent variable treats inputs that are not in the component as known (this can be done since those variables are evaluated earlier in the ordering on the DAG of components) and evaluates each variable in the component via a Gibbs sampling process. The values obtained upon convergence can then be passed to components later in the ordering. This follows by application of proposition 6 in \cite{lauritzen2002chain}, and so we can express the DAG factorization over chain components as:
	\begin{align*}
	p(\mathbf{V} \cup \mathbf{H}(\mathbf{f_A})) = \prod_{\mathbf{D} \in \mathcal{D}(\G_{\mathbf{f_A}}(\mathbf{V} \cup \mathbf{H}))} p(\mathbf{D} | \pa_{\G_{\mathbf{f_A}}(\mathbf{V} \cup \mathbf{H})}(\mathbf{D}))|_{\mathbf{A} = \mathbf{f_A}}\\
	\times \prod_{\mathbf{B} \in \mathcal{B}^{nt}(\G_{\mathbf{f_A}}(\mathbf{V} \cup \mathbf{H}))} p^{\star}(\mathbf{B} | \pa_{\G_{\mathbf{f_A}}(\mathbf{V} \cup \mathbf{H})}(\mathbf{B}))
	\end{align*}
	$\G_{\mathbf{f_A}}$ is a proper latent-variable chain graph.
	
	We derive the remainder of the proof via the argument in the proof of theorem 2 in \cite{sherman2018identification}. We assume without loss of generality that $\mathbf{Y}$ has no children in $\G(\mathbf{V})$.
	
	Consider the chain graph factorization of $\G_{\mathbf{f_A}}$ derived above. Because $\mathbf{H}$ is block-safe in $\G$, the non-trivial blocks term can be re-written as follows:
	\begin{align*}
		\prod_{\mathbf{B} \in \mathcal{B}^{nt}(\G_{\mathbf{f_A}}(\mathbf{V} \cup \mathbf{H}))} p^{\star}(\mathbf{B} | \pa_{\G_{\mathbf{f_A}}(\mathbf{V} \cup \mathbf{H})}(\mathbf{B})) &= \prod_{\mathbf{B} \in \mathcal{B}^{nt}(\G_{\mathbf{f_A}}(\mathbf{V}))} p(\mathbf{B} | \pa_{\G_{\mathbf{f_A}}(\mathbf{V})}(\mathbf{B}))\\
		&= \prod_{\mathbf{B} \in \mathcal{B}^{nt}(\tilde{\G}^b)} p^{\star}(\mathbf{B} | \pa_{\G_{\mathbf{f_A}}(\mathbf{V})}(\mathbf{B}))|_{\mathbf{A} = \tilde{\mathbf{a}}_{\mathbf{B}}}
	\end{align*}
	
	We are now left with the following factorization for the overall graph:
	\begin{align*}
		p(\{\mathbf{V} \cup \mathbf{H}\}&(\mathbf{f_A})) = \prod_{\mathbf{B} \in \mathcal{B}^{nt}(\tilde{\G}^b)} p^{\star}(\mathbf{B} | \pa_{\G_{\mathbf{f_A}}(\mathbf{V})}(\mathbf{B}))\\
		\times &\prod_{\mathbf{D} \in (\mathbf{V} \cup \mathbf{H}) \setminus \big(\bigcup_{\mathbf{B} \in \mathcal{B}^{nt}(\G_{\mathbf{f_A}})} \mathbf{B}\big)} \,\, \prod_{V \in \mathbf{D} \setminus \mathbf{A}} p(V | \pa_{\G_{\mathbf{f_A}}(\mathbf{V} \cup \mathbf{H})}(V)) \prod_{V \in \mathbf{D} \cap \mathbf{A}} f_V(\mathbf{Z}_V)|_{\mathbf{A} = \mathbf{f_A}}
	\end{align*}
	The factors in the second term are singleton nodes by construction and so they are defined by either observed $p(V | \pa_{\G_{\mathbf{f_A}}(\mathbf{V} \cup \mathbf{H})}(V))$ if $V \not \in \mathbf{A}$ and $f_V \in \mathbf{f_A}(\mathbf{Z}_V)$ if $V \in \mathbf{A}$.
	
	If we marginalize $\mathbf{H}$ from this second set of terms, using standard procedures \cite{tian2002general}, then the resulting expression is the kernel described in the statement of the theorem:
	$q(\mathbf{D}^{\star} | \pa^s_{\G(\mathbf{V})}(\mathbf{D}^{\star})) = \prod_{\mathbf{D} \in \mathcal{D}(\G_{\mathbf{f_A}})} q(\mathbf{D} | \pa^s_{\G_{\mathbf{f_A}}(\mathbf{V})}(\mathbf{D}))$, where $q(\mathbf{D} | \pa^s_{\G_{\mathbf{f_A}}(\mathbf{V})}(\mathbf{D})) = \prod_{D \in \mathbf{D}} p(\mathbf{D} | \mathbf{V}_{\prec \mathbf{D}})$ if $\mathbf{D} \cap \mathbf{A} = \emptyset$ and $q(\mathbf{D} | \pa^s_{\G_{\mathbf{f_A}}(\mathbf{V})}(\mathbf{D})) = f_A(\mathbf{Z}_A)$ if $\mathbf{D} \cap \mathbf{A} \neq \emptyset$.
	
	
	Since $\mathbf{Z_A}$ are all observed by assumption, we can manipulate this kernel as in the proof of soundness for theorem 2 in \cite{sherman2018identification}. Whereas in \cite{sherman2018identification} the authors fixed $\mathbf{A}$ to constants, here we can express setting $\mathbf{A}$ to stochastic values according to $\mathbf{f_A}$. The claim is then immediate.
	
	\textbf{Claim 2: The segregated graph policy ID formula is complete}
	
	We adapt the proof techniques in \cite{shpitser2018identification, sherman2018identification}. At a high level, we will use the fact that $p(\mathbf{Y}^{\star}(\mathbf{a}))$ is not identified to demonstrate that there is a hedge in $\G$. We will then extend the hedge down the graph to reach $\mathbf{Y}$ via $\ext_{\G_{\mathbf{Y}^{\star}}}(\text{hedge})$ and $\ant_{\G_{\mathbf{Y}^{\star}}}(\mathbf{Y})$ to show non-identification. We do this by arguing along the partially directed paths from the hedge to $\mathbf{Y}$, which requires considering subgraphs of $\G_{\mathbf{Y}^{\star}}$. We show non-identifiability in each of an increasingly restricted submodel of $\G_{\mathbf{Y}^{\star}}$ and then show that non-identification in the submodels yields non-identification in $\G_{\mathbf{Y}^{\star}}$. More concretely, there are two complications that must be dealt with for showing completeness of policy interventions: the hedge might intersect $\mathbf{Y}$ and we must extend the hedge down to $\mathbf{Y}$ via partially directed paths. We construct a subgraph for demonstrating the latter case and then a subgraph of that for the former case. We now proceed with the proof.
	
	Suppose $p(\mathbf{Y}^{\star}(\mathbf{a}))$ is not identified in $\G$. Then there is a district $\mathbf{D} \in \mathcal{D}(\G_{\mathbf{Y}^{\star}})$ that is not reachable in $\G$. Let $\mathbf{R} = \{D \in \mathbf{D} | \ch_{\G}(D) \cap \mathbf{D} = \emptyset\}$. Let $\mathbf{A}^{\star} = \mathbf{A} \cap \pa_{\G}(D)$. Then there exists $\mathbf{D}' \supset \mathbf{D}$, such that $\mathbf{D}$ and $\mathbf{D}'$ form a hedge for $p(\mathbf{R} | \text{do}(\mathbf{a}^{\star}))$ and thus $p(\mathbf{R} | \text{do}(\mathbf{a}^{\star}))$ is not identified by \cite{shpitser2006identification}.
	
	Let $\mathbf{Y}'$ be the minimal subset of $\mathbf{Y}$ such that $\mathbf{R} \subseteq \ant_{\G_{\mathbf{f_A}}}(\mathbf{Y}')$. Consider a subgraph $\G^{\dagger}$ of $\G_{\mathbf{f_A}}$, with vertices $\mathbf{V}' \subseteq \mathbf{V}$, consisting of all edges in $\G$ in the hedge on $\mathbf{D}, \mathbf{D}'$ described above, and edges that lie in partially directed paths in $\G_{\mathbf{f_A}}$ from $\mathbf{R}$ to $\mathbf{Y}'$. We restrict attention, without loss of generality, to at most one child per node in each partially directed path such that our paths form a forest from $\mathbf{R}$ to $\mathbf{Y}'$. By Lemma \ref{lem:g-fa}, $\G^{\dagger}$ does not contain any directed, nor partially directed cycles. Let $\mathbf{A}^{\dagger} = \{\mathbf{A}^{\star} \cup A | A \in \mathbf{A} \text{ in } \G^{\dagger} \}$. For each $A^{\dagger} \in \mathbf{A}^{\dagger}$, we restrict attention to policies that map from $\mathbf{Z}_{A^{\dagger}}^{\dagger}$ to $A^{\dagger}$, where $\mathbf{Z}_{A^{\dagger}}^{\dagger} = \mathbf{Z}_{A^{\dagger}} \cap \mathbf{V}'$.
	
	Now, following the proof of theorem 2 in the supplement of \cite{sherman2018identification}, we define an ADMG $\tilde{\G}^{\dagger}$ which has the same vertices and edges as the $\mathbf{D}, \mathbf{D}'$ hedge in $\G^{\dagger}$, and has a copy of each vertex in each partially directed path from $\mathbf{R}$ to $\mathbf{Y}'$ in $\G^{\dagger}$ but replaces all the undirected edges on those partially directed paths with directed edges oriented away from $\mathbf{R}$ towards $\mathbf{Y}'$. We denote the variable copies in $\tilde{\G}^{\dagger}$ corresponding to $\mathbf{Y}'$ in $\G^{\dagger}$ by $\tilde{\mathbf{Y}}'$. This orientation is possible because each undirected edge either corresponds to a (known) policy in the intervention set, or to an observed structural equation. In either case, the observed distribution continues to argree between the two counterexamples witnessing non-identifiability. For $\mathbf{A}^{\dagger}$ in $\tilde{\G}^{\dagger}$, we further restrict attention to policies inducing directed edges from $\mathbf{R}$ to $\tilde{\mathbf{Y}}'$ (i.e. ignoring policies going the opposite direction that induce undirected edges). We denote these nodes by $\tilde{\mathbf{A}}^{\dagger}$.
	
	We now show that $p(\tilde{\mathbf{Y}}'(\{\tilde{A}^{\dagger} = f_{\tilde{A}^{\dagger}} | \tilde{A}^{\dagger} \in \tilde{\mathbf{A}}^{\dagger} \}))$ is not identified in $\tilde{\G}^{\dagger}$ following the argument in the proof of theorem 6 in the supplement of \cite{shpitser2018identification}.
	Observe that for $\mathbf{R} \subseteq \tilde{\mathbf{Y}}'$, the subclaim is immediate by the recursive argument in the proof of theorem 4 in \cite{shpitser2018identification}. Otherwise, pick a node $\tilde{Y}'$ in $\tilde{\G}^{\dagger}$ such that $\pa_{\tilde{\G}^{\dagger}}(\tilde{Y}') \subseteq \mathbf{R}$ and $\pa_{\tilde{\G}^{\dagger}}(\tilde{Y}') \setminus \tilde{\mathbf{Y}}' \neq \emptyset$ (as in \cite{shpitser2018identification}, such a vertex must exist since $\tilde{\G}^{\dagger}$ is acyclic and $\mathbf{R} \setminus \tilde{\mathbf{Y}}' \neq \emptyset$). If this $\tilde{Y}' \in \tilde{\mathbf{A}}^{\dagger} \setminus \mathbf{A}^{\star}$, the subclaim is immediate since $\tilde{Y}'$ does not intersect our hedge and we can extend down the graph using the argument in theorem 4 of \cite{shpitser2018identification}.
	
	If $\tilde{Y}' \in \mathbf{A}^{\star}$ then we can create a graph $\bar{\G}$ by copying the variables on the path $\tilde{Y}' \rightarrow V_1 \rightarrow \dots \rightarrow \bar{Y} \in \tilde{\mathbf{Y}}'$ in $\tilde{\G}^{\dagger}$. We then apply the argument in theorem 4 of \cite{shpitser2018identification} to show that $p(\bar{Y}(\mathbf{a}^{\star}))$ is not identified along this path when we set $\mathbf{a}^{\star}$ according to the policies specified by $\mathbf{f_{A^{\star}}}$. This follows since, by assumption, $\mathbf{f_{A^{\star}}} \subseteq \mathbf{f_A}$ lies in an unrestricted policy class. Now, as $p(\bar{Y}(\mathbf{a}^{\star}))$ is not identified in $\bar{\G}$, we can use the two counterexamples witnessing non-identifiability in $\bar{\G}$ to obtain non-identifiability for $p(\tilde{\mathbf{Y}}'(\mathbf{f_{\tilde{A}^{\dagger}}}))$. To do so, we define new variables in $\tilde{\G}^{\dagger}$ that are the Cartesian product of variable copies created in $\bar{\G}$ and their corresponding variables in $\tilde{\G}^{\dagger}$. Non-identifiability follows via the standard argument in lemma 1 of \cite{shpitser2018identification}.
	
	Now that we have shown that $p(\tilde{\mathbf{Y}}'(\{\tilde{A}^{\dagger} = f_{\tilde{A}^{\dagger}} | \tilde{A}^{\dagger} \in \tilde{\mathbf{A}}^{\dagger} \}))$, we have two counterexamples witnessing non-identifiability in $\tilde{\G}^{\dagger}$ which agree on the observed data distribution but disagree on the counterfactual distribution. We use these counterexamples to demonstrate non-identifiability of $p(\mathbf{Y}'(\{A = f_A | A \in \mathbf{A}^{\star}\}))$ in $\G^{\dagger}$. To do so, we define variables along the partially directed paths from $\mathbf{R}$ to $\mathbf{Y}'$ in $\G^{\dagger}$. These variables are created by taking the Cartesian product of variable copies in $\tilde{\G}^{\dagger}$ and the corresponding variables in $\G^{\dagger}$. As before, the counterexamples continue to agree on the observed data distribution and disagree on the counterfactual distribution. Thus $p(\mathbf{Y}'(\{A = f_A | A \in \mathbf{A}^{\star}\}))$ is not identified in $\G^{\dagger}$. Since $\mathbf{Y}' \subseteq \mathbf{Y}^{\star}$, the result is immediate, subject to the remaining argument on the chain graph properties of $\G^{\dagger}$ and $\tilde{\G}^{\dagger}$ below.
	
	Following \cite{sherman2018identification}, fix a block $\mathbf{B}$ in $\G^{\dagger}$. For any $B \in \mathbf{B}$, there exists a set of variables $B_1, \dots, B_k$ in $\tilde{\G}^{\dagger}$ such that $B$ is defined as the Cartesian product of $B_1, \dots, B_k$. Any variable $A \in \nb_{\G^{\dagger}} \cup \pa_{\G^{\dagger}}(B)$ is similarly a Cartesian product of $A$ variables. Then it follows that $B \ci ((\pa_{\G^{\dagger}}(\mathbf{B}) \cup \mathbf{B}) \setminus (\nb_{\G^{\dagger}}(B) \cup \pa_{\G^{\dagger}}(B))) | (\nb_{\G^{\dagger}}(B) \cup \pa_{\G^{\dagger}}(B))$ by d-separation rules in the ADMG $\tilde{\G}^{\dagger}$ and that there are no colliders in $\tilde{\G}^{\dagger}$. These both follow from our vertex copy argument which separates out $B$ from the rest of the block and eliminates the possibility of colliders by making every path from $\mathbf{R}$ to $\mathbf{Y}'$ a partially directed chain. This demonstrates that $\G^{\dagger}$ and $\tilde{\G}^{\dagger}$ (and trivially $\bar{\G}$) satisfy the independence constraints implied by the CG Markov property, thus proving the claim.
\end{prf}

\section{Derivation of the Figure \ref{fig:gross-sg} Functional}
From Fig. \ref{fig:gross-sg}(a), we obtain $\G_{\mathbf{f_A}}$ in Fig. \ref{fig:gross-sg}(b) by applying the intervention detailed in Table \ref{tab:intervention}. In turn, from this post-intervention graph we observe that $\mathbf{Y}^{\star} = \ant_{\G_{\mathbf{f_A}}}(\mathbf{Y}) \setminus \mathbf{A} = \{C_2, C_3, M_3, Y_2, Y_3\}$ and obtain the induced subgraph $\G_{\mathbf{Y}^{\star}}$ in Fig. \ref{fig:gross-sg}(c).

$\G_{\mathbf{Y}^{\star}}$ factorizes into kernels relating to district nodes and block nodes: $q_{\mathcal{D}}(C_1, A_1, M_1, Y_1, Y_2, Y_3|C_2, M_2, M_3)$ and $q_{\mathcal{B}}(M_2, M_3, A_2, A_3, C_2, C_3 | \emptyset)$. The block nodes factorize as a product of blocks, as in the first term of Eq. \ref{eq:sg-policy-id-apx}:
\begin{align*}
	q_{\mathcal{B}}(\mathbf{B}^{\star}|\pa_{\G_{\mathbf{f_A}}}(\mathbf{B}^{\star})) &= \prod_{\mathbf{B} \in \mathcal{B}(\tilde{\G}^{b})} p^{\star}(\mathbf{B} | \pa_{\G_{\mathbf{f_A}}}(\mathbf{B}))\\
	&= p^{\star}(M_2, M_3 | A_2, A_3, C_2) p^{\star}(A_2, A_3 | C_2, C_3) p^{\star}(C_2, C_3)
\end{align*}
Note that $p^{\star}(C_2, C_3) = p(C_2, C_3)$ since the $C_2 - C_3$ block is unchanged relative to the observed data.

Separately, we must fix sets for each $\G_{\mathbf{Y}^{\star}}$ district $\{\{M_3\}, \{Y_2, Y_3\}\}$ in $q_{\mathcal{D}(\G)}$. The derivations of these pieces is as follows:
\begin{align*}
	\phi_{\mathbf{D}^{\star} \setminus \{M_3\}}(q(C_1, A_1, M_1, Y_1, Y_2, Y_3|C_2, M_2, M_3); \G^d) = \phi_{\mathbf{D}^{\star}}(q(C_1, A_1, M_1, Y_1, Y_2, Y_3|C_2, M_2, M_3); \G^d)
\end{align*}
This follows since $M_3$ is already fixed in this kernel and subgraph. Since we must fix all variables in the kernel and all variables in the kernel are fixable, this term simplifies to $p(\emptyset) = 1$.

For the second kernel, we have: $\phi_{\mathbf{D}^{\star} \setminus \{Y_2, Y_3\}}(q(C_1, A_1, M_1, Y_1, Y_2, Y_3|C_2, M_2, M_3); \G^d)$
\begin{align*}
	&= \phi_{C_1, A_1, M_1, Y_1}(q(C_1, A_1, M_1, Y_1, Y_2, Y_3|C_2, M_2, M_3); \G^d)\\
	&= \phi_{A_1, M_1, Y_1}(\frac{q(C_1, A_1, M_1, Y_1, Y_2, Y_3|C_2, M_2, M_3)}{p(C_1)}; \phi_C(\G^d))\\
	&= \phi_{A_1, M_1, Y_1}(q(A_1, M_1, Y_1, Y_2, Y_3|C_2, M_2, M_3, C_1); \phi_{C_1}(\G^d))\\
	&= \phi_{M_1, Y_1}(\frac{q(A_1, M_1, Y_1, Y_2, Y_3|C_2, M_2, M_3, C_1)}{p(A_1 | C_1)}; \phi_{C_1, A_1}(\G^d))\\
	&= \phi_{M_1, Y_1}(q(M_1, Y_1, Y_2, Y_3|C_2, M_2, M_3, C_1, A_1); \phi_{C_1, A_1}(\G^d))\\
	&= \phi_{Y_1}(\frac{q(M_1, Y_1, Y_2, Y_3|C_2, M_2, M_3, C_1, A_1)}{p(M_1 | A_1)}; \phi_{C_1, A_1, M_1}(\G^d))\\
	&= \phi_{Y_1}(q(Y_1, Y_2, Y_3|C_2, M_2, M_3, C_1, A_1, M_1); \phi_{C_1, A_1, M_1}(\G^d))\\
	&= \frac{q(Y_1, Y_2, Y_3|C_2, M_2, M_3, C_1, A_1, M_1)}{p(Y_1 | A_1, M_1)}; \phi_{C_1, A_1, M_1, Y_1}(\G^d))\\
	&= p(Y_2, Y_3 | C_1, C_2, M_1, M_2, M_3, A_1, Y_1)
\end{align*}

This yields the functional for $p(\{Y_2, Y_3\}(\mathbf{f_A}))$:
	\begin{align*}
	\sum_{\{A_1, A_2, A_3, M_2, M_3, C_2, C_3\}} &\Big(p^{\star}(M_2, M_3 | A_2, A_3, C_2)p^{\star}(A_2, A_3 | C_2, C_3) p^{\star}(C_2, C_3)\\
	&\times p(Y_2, Y_3 | C_1, C_2, M_1, M_2, M_3, A_1, Y_1)\Big)
	\end{align*}

\section{Experimental Details and Extended Results}
Each $C_i, A_i, Y_i$ are generated according to the following densities (note that $C_i$ is a 3-dimensional vector):
\begin{align*}
	C_{i, j} &\sim \text{Beta}(\alpha_j, \beta_j)\\
	p(A_i = 1 | C_i, C_{-i}) &= \text{expit}(\sum_{j = 1}^{3} \gamma_j C_{i, j} + \frac{\tau_{AC}}{|\mathcal{N}_i|} \sum_{k \in \mathcal{N}_i} \sum_{j = 1}^{3} C_{k, j})\\
	p(Y_i = 1 | A_i, A_{-i}, C_i, C_{-i}, Y_{-i}) &= \text{expit}\bigg(\eta A_i + \sum_{j = 1}^{3} \delta_j C_{i, j}\\
	&+ \frac{1}{|\mathcal{N}_i|} \sum_{k \in \mathcal{N}_i} \Big(\tau_{YA} A_k + \tau_{YY} Y_k + \sum_{j = 1}^{3} \tau_{YC} C_{k, j}\Big)\bigg)
\end{align*}
where $\mathcal{N}_i$ denote unit $i$'s neighbors in $\G_{\mathbf{f_A}}$.

The parameters for the Beta distribution for $C$ for both types of experiments (policy and bias) are given by:
\begin{table}[h]
\begin{center}
	\begin{tabular}{|l|l|}
		\hline
		$\alpha$ & $\beta$ \\ \hline
		1.5 & 3 \\ \hline
		6 & 2 \\ \hline
		.8 & .8 \\ \hline
	\end{tabular}
\end{center}
\caption{Parameters for generating $C_i$}
\end{table}

The parameters for $A_i$ and $Y_i$ differ between the bias and policy experiments. For $A$ we have:
\begin{table}[h]
\begin{center}
	\begin{tabular}{|l|l|l|}
		\hline
		Parameter & Bias & Policy \\ \hline
		$\gamma_1$ & 1 & .5 \\ \hline
		$\gamma_2$ & 0 & .2 \\ \hline
		$\gamma_3$ & 0 & .25 \\ \hline
		$\tau_{AC}$ & 0 & .15 \\ \hline
	\end{tabular}
\end{center}
\caption{Parameters for generating $A_i$}
\end{table}

And for $Y$ we have:
\begin{table}[h]
\begin{center}
	\begin{tabular}{|l|l|l|}
		\hline
		Parameter & Bias & Policy \\ \hline
		$\eta$ & -3 & .6 \\ \hline
		$\delta_1$ & 1 & -.3 \\ \hline
		$\delta_2$ & 0 & .4 \\ \hline
		$\delta_3$ & 0 & .1 \\ \hline
		$\tau_{YA}$ & 3 & .2 \\ \hline
		$\tau_{YY}$ & .1 & .3 \\ \hline
		$\tau_{YC}$ & 0 & -.2 \\ \hline
	\end{tabular}
\end{center}
\caption{Parameters for generating $Y_i$}
\end{table}

Finally, for the policy experiment we have results similar to those in the main draft, which demonstrate the efficacy of policy interventions in selection actions that yield a more optimal outcome.
\begin{figure*}[h]
	\begin{center}
		\subfloat[]{
			\includegraphics[width=2.5in]{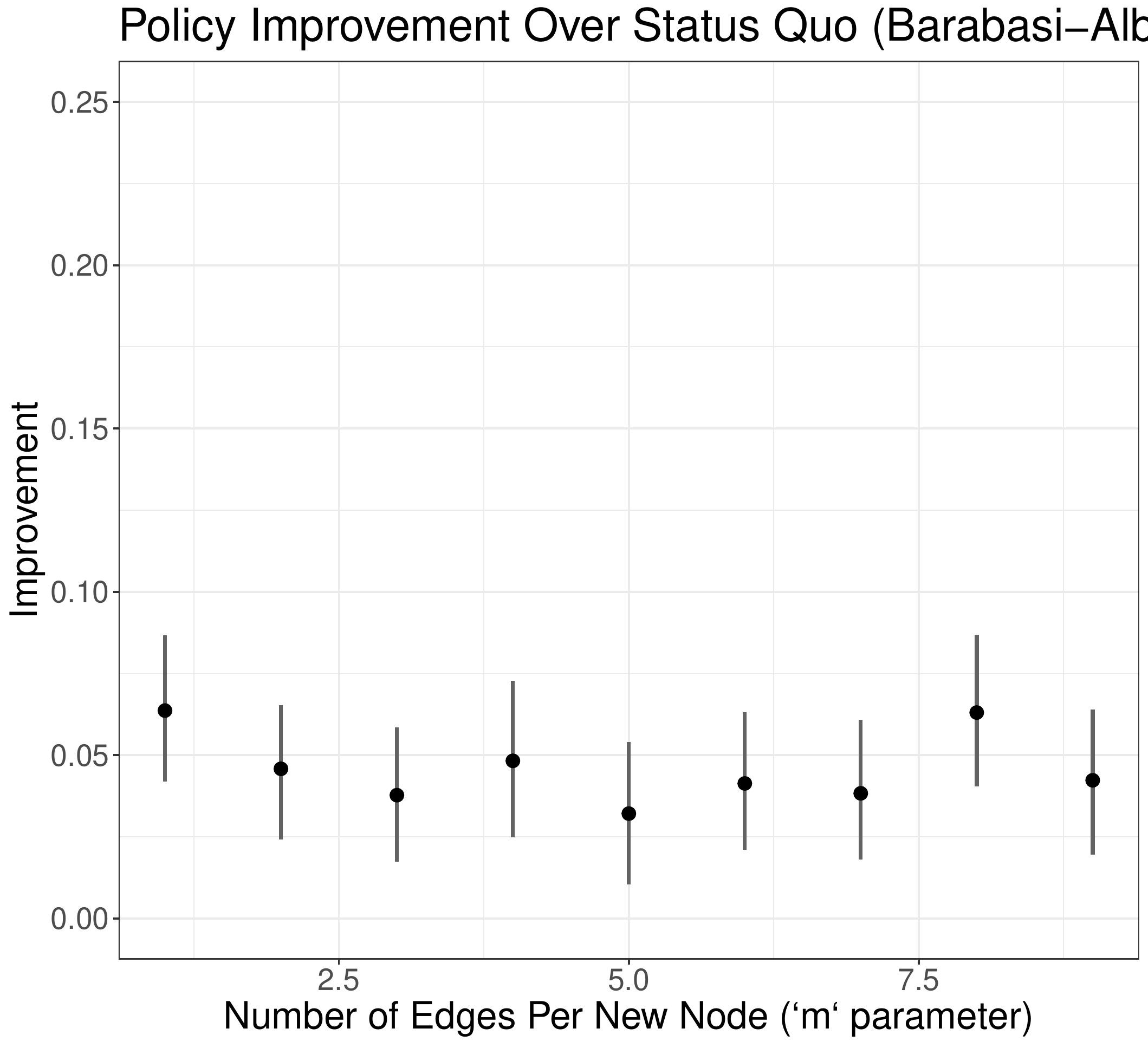}
			\label{fig:barabasi}
		}
		\hspace{.5cm}
		\subfloat[]{\includegraphics[width=2.5in]{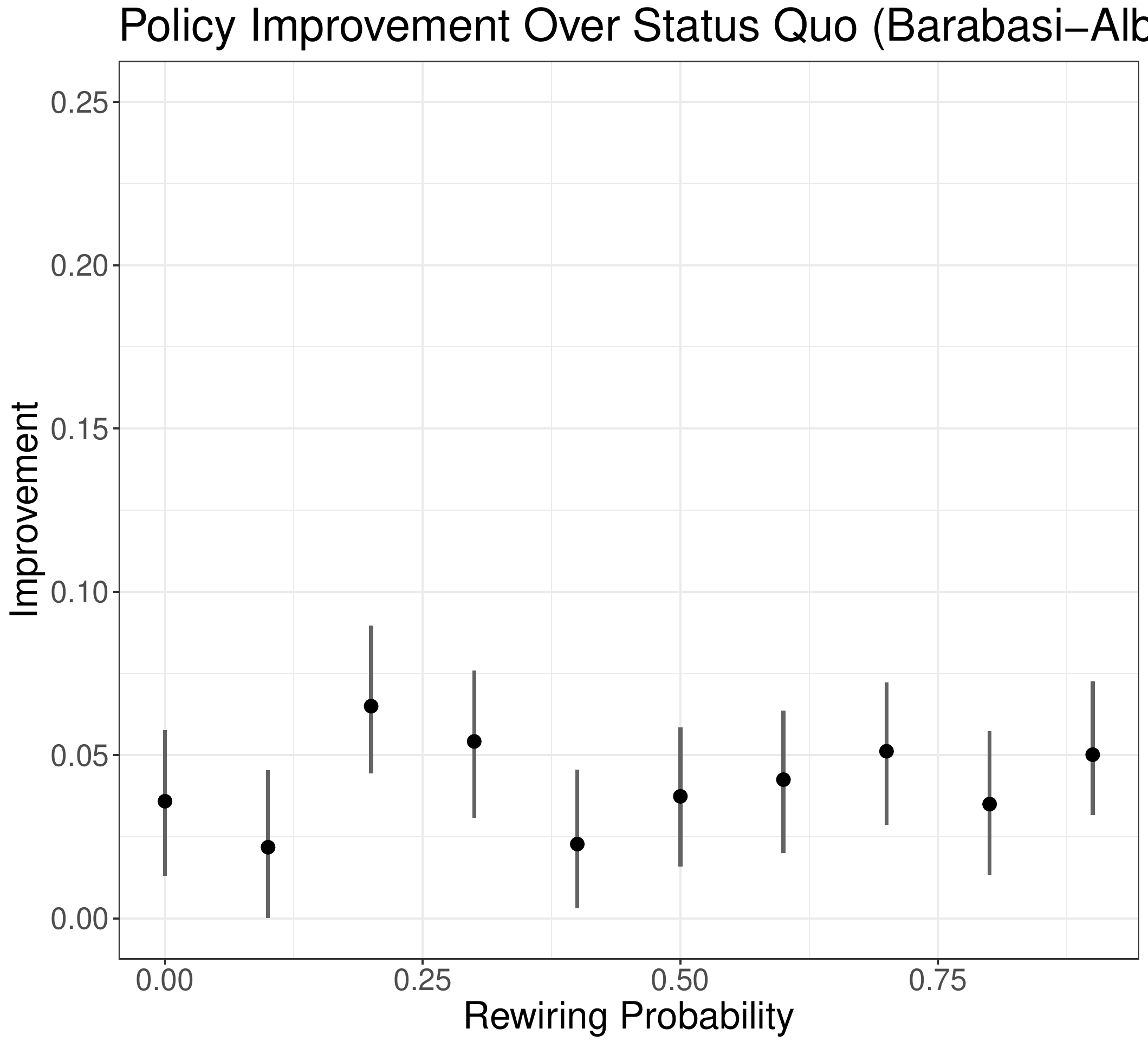} \label{fig:watts}}
		\caption{Difference in expected outcomes between adopting an optimal strategy and using the status quo strategy in the Barab\'asi-Albert model \ref{fig:barabasi} and the Watts-Strogatz small world model \ref{fig:watts}. We perform these analyses at several network densities to demonstrate the general efficacy of this approach.}
	\end{center}
\end{figure*}

\clearpage

\bibliographystyle{abbrv}
\bibliography{references}

\end{document}